\documentclass[12pt]{article}

\usepackage{wrapfig}
\usepackage{graphicx}
\usepackage[none]{hyphenat}
\usepackage[english]{babel}
\usepackage[utf8]{inputenc}
\usepackage[T1]{fontenc}
\usepackage{parskip}

\usepackage{authblk}
\usepackage{tikz-cd}

\usepackage{caption}
\usepackage{csquotes}
\usepackage{mathtools,amssymb,amsfonts,amsthm}

\usepackage[breaklinks,unicode=true]{hyperref}
\usepackage[capitalise]{cleveref} 

\theoremstyle{plain}
\newtheorem{theorem}{Theorem}
\newtheorem*{theorem*}{Theorem}

\newtheorem*{lemma*}{Lemma}
\newtheorem{proposition}[theorem]{Proposition}
\newtheorem*{proposition*}{Proposition}
\newtheorem{corollary}[theorem]{Corollary}
\newtheorem*{corollary*}{Corollary}

\theoremstyle{definition}
\newtheorem{definition}{Definition}
\newtheorem*{definition*}{Definition}
\newtheorem{example}{Example}
\newtheorem*{example*}{Example}

\crefname{theorem}{Theorem}{Theorems}
\Crefname{theorem}{Theorem}{Theorems}
\crefname{lemma}{Lemma}{Lemmas}
\Crefname{lemma}{Lemma}{Lemmas}
\crefname{proposition}{Proposition}{Propositions}
\Crefname{Prop}{Proposition}{Propositions}
\crefname{corollary}{Corollary}{Corollaries}
\Crefname{corollary}{Corollary}{Corollaries}
\crefname{definition}{Definition}{Definitions}
\Crefname{definition}{Definition}{Definitions}
\crefname{example}{Example}{Examples}
\Crefname{example}{Example}{Examples}
\crefname{theorem}{Theorem}{Theorems}

\newtheorem*{exercise*}{Exercise}
\crefname{exercise}{exercise}{exercises}
\Crefname{exercise}{Exercise}{Exercises}  

\theoremstyle{remark}
\newtheorem{remarkx}{Remark}
\newtheorem*{remarkx*}{Remark}
\crefname{remark}{Remark}{Remarks}
\Crefname{remark}{Remark}{Remarks}
\newenvironment{remark}
  {\pushQED{\qed}\remarkx}
  {\popQED\endremarkx}
\newenvironment{remark*}
  {\pushQED{\qed}\remarkx*}
  {\popQED\endremarkx*}

\title{A geometric model for non-uniform processes of morphogenesis}

\author[-]{V\'ictor Manuel Jim\'enez}
\author[+*]{Manuel de León}

\affil[1]{\href{mailto:victor.jimenez@mat.uned.es}{victor.jimenez@mat.uned.es}}

\affil[2]{\href{mailto:mdeleon@icmat.es}{mdeleon@icmat.es}}

\affil[-]{Universidad Nacional de Educación a Distancia (UNED), Departamento de Matem\'aticas Fundamentales.
Calle de Juan del Rosal 10, 28040, Madrid, Spain} 

\affil[*]{Instituto de Ciencias Matem\'aticas (CSIC-UAM-UC3M-UCM),
C\textbackslash Nicol\'as Cabrera, 13-15, Campus Cantoblanco, UAM
28049 Madrid, Spain} 

\affil[+]{Real Academia de Ciencias Exactas, Fisicas y Naturales, C/de Valverde
22, 28004 Madrid, Spain}

\date{\today}

\begin{document}

\sloppy

\maketitle

\begin{abstract}

In this paper we present an application of the groupoid theory to the study of relevant case of material evolution phenomena, the \textit{process of morphogenesis}. Our theory is inspired by Walter Noll's theories of continuous distributions and provides a unifying and very simple framework of these phenomena. We present the explicit equation, the \textit{morphogenesis equation}, to calculate the \textit{material distributions} associated to this phenomenon.
\end{abstract}

\tableofcontents

\section{Introduction}

Our approach to the mechanics of continuous media follows Walter Noll's theory, in which a constitutive law determines the mechanical behavior of the material under deformation \cite{WNOLLTHE,WNOLL}. The notion of groupoid is intimately linked to the study of materials, and in fact, in the works of Walter Noll one can implicitly recognize the existence of this algebraic structure when considering the collection of all material isomorphisms obtained through a given constitutive law.

For a simple material $\mathcal{B}$, the collection of the material isomorphisms constitutes the so-called material groupoid, $\Omega \left( \mathcal{B}\right)$, which is in fact a subgroupoid of the Lie groupoid formed by all the linear isomorphisms between the tangent spaces of all the pairs of points of the material body \cite{MATGROUPALG,VMJIMM}. This is the so-called \textit{material groupoid} and it is not necessarily a Lie groupoid. This lack of differentiability measures the lack of differentiable uniformity of the material (for explicit examples see \cite{MD,GENHOM}).

On the other hand, a natural question is what infinitesimal properties of the material groupoid reflect inhomogeneity. If it is a Lie subgroupoid, we can associate with it the corresponding Lie algebroid and characterize the homogeneity by its properties (in short, it is a global way of dealing with material G-structures that had been considered in other approaches \cite{MAREMDL7,MAREMDL6,MELZA,MELZASEG}). But even if the material groupoid is not Lie, we have been able to construct a generalization of the associated Lie algebroid, which we call \textit{characteristic distribution} and which is still able to give answers to the concepts of homogeneity and integrability \cite{VMMDME}.

This distribution is integrable (in the sense of Stefan and Sussmann \cite{PS,HJS}) and determines generalized foliations, one in the subgroupoid and the other in the body manifold, so that, roughly speaking, differentiability is introduced where there was none, and the leaves in the material groupoid become true Lie groupoids on the leaves in the base on which they project.

We are now interested in a new concept, the possible evolution of the material. To address this issue, we need to consider constitutive laws that take into account time (or any other evolutionary parameter(s)), so that the material body $\mathcal{B}$ is replaced by the \textit{space-time fibre bundle} $\mathcal{C} = \mathbb{R} \times \mathcal{B} \longrightarrow \mathbb{R}$, and more specifically, by the vertical fibre of this fibration, which is a vector subbundle of the tangent bundle $T \mathcal{C}$. $\mathcal{C}$ is also called the \textit{body-time manifold}. So, we can introduce the notion of \textit{time-material isomorphism}, and the corresponding
material groupoid $\Omega \left( \mathcal{C}\right)$.

As opposed to the uniformity in the spatial case, new material properties arise, associated with the evolution of the body. In particular, the temporal counterpart of uniformity is a specific case of evolution of the material called \textit{remodeling}. Intuitively, a material evolution presents a remodeling when the constitutive properties of the material does not change with the time. This kind of evolution may be found in biological tissues \cite{RODRIGUEZ1994455}. Wolff’s law of trabecular architecture of bones is another relevant example \cite{TURNER19921}. Growth and resorption are given by a remodeling with volume increase or volume decrease of the material body $\mathcal{B}$.

We say that $\mathcal{C}$ is a process of \textit{aging} if it is not a process of remodeling. Clearly, if the material response is not preserved along the time via material isomorphisms, then the constitutive properties are changing with the time. In \cite{EvoEqu, CharFol}, we use the corresponding characteristic distributions and their foliations to present several results characterizing the different types of remodeling and aging. Here, however, we are interested in a particular case of evolution: \textit{morphogenesis}.

In order to work with the concept of morphogenesis, we will need to extend this notion to groupoids, introducing the notion of \textit{normalizoid} of a subgroupoid in a groupoid. This generalization may be found in \cite{CEDRIC12} to study uniformity and homogeneity in functional graded materials (FGM).

A material point $X$ is said to undergo a process of evolution without morphogenesis when its symmetry groups at all the instants are conjugated. $\mathcal{C}$ is said to be a process of evolution without morphogenesis when all the points undergo a process of evolution without morphogenesis. Hence, remodeling is an example of the process of evolution without morphogenesis. In other words, as it is natural, any process of morphogenesis is a particular case of aging. Thus, a process of morphogenesis entails a breakdown of symmetry, a sudden change in the material symmetry type. Using the technique of characteristic distributions, we have been able to express it in terms of a differential equation, which we call the morphogenesis equation.

The paper is structured as follows. In section \ref{chardist345} we present an overview of the definitions and results needed in groupoid theory and their associated characteristic distributions. In section \ref{partelasticmat} we introduce the concept of material evolution of simple bodies, while in section \ref{section4} we construct the material groupoid associated with the vertical body-time manifold, next to the associated material distributions. Section \ref{section5} is devoted to discuss the phenomenon of morphogenesis. It is convenient to point out that the main development of this paper is contained in this section. First, we start studying a natural generalization of the notion ``\textit{normalizer}'' and ``\textit{normal subgroup}'' from groups to groupoids. Here, we prove some general results (Proposition \ref{propositionaux1}, Proposition \ref{2.2} and Corollary \ref{corollaryaux1}), but we do not do a deep study of this theory, because our interest is focused on its application to Continuum Mechanics. Then, we use this notion to deal with the notion of morphogenesis. In fact, not only theoretical constructions are made, but also the linear equations that determine the morphogenesis are obtained. In particular, it is important to highlight that, we give a specific way to construct the material distribution which characterize the phenomenon of morphogenesis. Namely, the material distribution is pointwise generated by the left-invariant vector fields $\Theta$ on $\Phi \left( \mathcal{V} \right)$, the space of linear isomorphisms between fibres of the vertical bundle $\mathcal{V}$ of the body-time manifold, satisfying that
\begin{equation*}
 TW \left( \left[ \Theta , \Lambda \right] \right) = 0,
 \end{equation*}
for any left invariant admissible vector field $\Lambda$ for the couple $ \left( \Phi \left( \mathcal{V}\right),  \Omega \left( \mathcal{C} \right) \right)$ which is tangent to the $\alpha-$fibres. This fact permits us to introduce the \textit{morphogenesis equations},
\begin{equation}
y^{i}_{l}\left( \Theta^{l}_{r}\Lambda^{r}_{j}  + \Theta^{k}\dfrac{\partial\Lambda^{l}_{j}}{\partial x^{k}}    +  \lambda \dfrac{\partial \Lambda_{j}^{l}}{\partial t}    -    \Lambda^{l}_{r}\Theta^{r}_{j} \right) \dfrac{\partial   W}{\partial y^{i}_{j} } = 0
\end{equation}
for all matrix function $\left(\Lambda^{i}_{j}\right)$ fulfilling the equation,
\begin{equation*}
y^{i}_{l} \Lambda^{l}_{j}\dfrac{\partial   W}{\partial  y^{i}_{j} } = 0
\end{equation*}
Here the solutions are the functions $\lambda$, $\Theta^{i}$, and $\Theta^{j}_{i}$. We present analogous constructions for the evolution of a particle $X$. The above distribution, and its associated equation, are used to characterize the processes of morphogenesis in such a way that the dimension of the space of solutions will provide us the information about the existence of processes of morphogenesis in the evolution of the material.

\section{An onverview on groupoids and distribution}\label{chardist345}
We will give here a very brief introduction on \textit{(Lie) groupoids} and the relation with \textit{(smooth) distributions} which is crucial to understand the results proved in this paper. For a detailed study we refer to \cite{VMMDME} (see also \cite{CHARDIST,MD}).\\

On the one hand, groupoids may be though as a natural generalization of the notion of group, and they were introduced in 1926 by Brandt \cite{HBU}. Adding differential structure, we have the notion of \textit{Lie groupoid} which is due to Ehresmann (\cite{CELC,CELP,CES,CEC}) and Pradines (\cite{JPRA}). We will follow the reference on groupoids \cite{KMG} (see also \cite{EPSBOOK}, \cite{WEINSGROUP} or \cite{JNM} for a more more intuitive view).

\begin{definition}
\rm
Let $ M$ be a set. A \textit{groupoid} over $M$ is given by a set $\Gamma$ equipped with the maps $\alpha,\beta : \Gamma \rightarrow M$ (\textit{source map} and \textit{target map} respectively), $\epsilon: M \rightarrow \Gamma$ (\textit{section of identities}), $i: \Gamma \rightarrow \Gamma$ (\textit{inversion map}) and $\cdot : \Gamma_{\left(2\right)} \rightarrow \Gamma$ (\textit{composition law}). Here, $\Gamma_{(k)}$ denotes the $k$-tuplas $ \left(g_{1}, \hdots , g_{k}\right) \in \Gamma \times \stackrel{k)}{\ldots} \times \Gamma $ such that $\alpha\left(g_{i}\right)=\beta\left(g_{i+1}\right)$ for $i=1, \hdots , k -1$. The following properties are satisfied:\\
\begin{itemize}
\item[(1)] $\alpha$ and $\beta$ are surjective and for each $\left(g,h\right) \in \Gamma_{\left(2\right)}$,
$$ \alpha\left(g \cdot h \right)= \alpha\left(h\right), \ \ \ \beta\left(g \cdot h \right) = \beta\left(g\right).$$
\item[(2)] Associativity of the composition law, i.e.,
$$ g \cdot \left(h \cdot k\right) = \left(g \cdot h \right) \cdot k, \ \forall \left(g,h,k\right) \in \Gamma_{\left(3\right)}.$$
\item[(3)] For all $ g \in \Gamma$,
$$ g \cdot \epsilon \left( \alpha\left(g\right)\right) = g = \epsilon \left(\beta \left(g\right)\right)\cdot g .$$
In particular,
$$ \alpha \circ  \epsilon \circ \alpha = \alpha , \ \ \ \beta \circ \epsilon \circ \beta = \beta.$$
\item[(4)] For each $g \in \Gamma$,
$$i\left(g\right) \cdot g = \epsilon \left(\alpha\left(g\right)\right) , \ \ \ g \cdot i\left(g\right) = \epsilon \left(\beta\left(g\right)\right).$$
Then,
$$ \alpha \circ i = \beta , \ \ \ \beta \circ i = \alpha.$$
\end{itemize}
These maps ($\alpha$, $\beta$, $\epsilon$, $i$, and $\cdot$) will be called the \textit{structure maps}. We will denote this groupoid by $ \Gamma \rightrightarrows M$.
\end{definition}
Observe that, since $\alpha$ and $\beta$ are surjective we get
$$ \alpha \circ \epsilon = Id_{M}, \ \ \ \beta \circ \epsilon = Id_{M},$$
where $Id_{M}$ is the identity at $M$.\\
\noindent{Sometimes $M$ is denoted by $\Gamma_{\left(0\right)}$ and it is identified with the set $\epsilon \left(M\right)$ of identities of $\Gamma$. $\Gamma$ is also denoted by $\Gamma_{\left(1\right)}$. The elements of $M$ are called \textit{objects} and the elements of $\Gamma$ are called \textit{morphishms}. The map $\left(\alpha , \beta\right) : \Gamma \rightarrow M \times M$ is called the \textit{anchor map} and the space of sections of the anchor map is denoted by $\Gamma_{\left(\alpha, \beta\right)} \left(\Gamma\right)$. Finally, for each $g \in \Gamma$ the element $i \left( g \right)$ is denoted by $g^{-1}$.}\\
One may think that the definition of groupoid looks too ``\textit{abstract}''. However, roughly speaking, a groupoid may be depicted as a set of ``\textit{arrows}'' ($\Gamma$) joining points ($M$), in such a way that any two arrows may composed if the ending point of one coincides with the starting point of the other. Then, assuming natural conditions derived of the properties of a composition in a group, we get the definition of groupoid.

\begin{definition}
\rm

A \textit{subgroupoid} of a groupoid $\Gamma \rightrightarrows M$ is a groupoid $\Gamma' \rightrightarrows M'$ such that $M' \subseteq M$, $\Gamma' \subseteq \Gamma$ and the the structure maps of $\Gamma'$ are the restrictions of the structure maps of $\Gamma$.
\end{definition}
In particular, a subgroupoid has the same composition law of the correspondent groupoid.

\begin{example}\label{5}
\rm
A group $G$ is a groupoid over a point and the operation law of the groupoid, $\cdot$, is the operation in $G$.
\end{example}
\begin{example}\label{7}

\rm
For any set $X$, the product space $ X \times X$ is a groupoid over $X$, called the \textit{pair groupoid}, in such a way that,
\begin{itemize}
\item[] $\alpha \left(a,b\right) = a , \ \ \beta \left(a,b\right)=b, \ \forall \left(a,b\right) \in  X \times X$
\item[] $\left(c,b\right)\cdot\left(a,c\right)= \left(a,b\right), \ \forall \left( c,b\right),\left(a,c\right) \in X\times X$
\end{itemize}

\begin{example}\label{truvialG}
\rm
Let $X$ be a set and $G$ be a group. Then we can construct a groupoid $ X \times X \times G\rightrightarrows X$ where the source map is the second projection, the target map is the third projection and the composition law is given by the composition in $G$, i.e.,
$$\left( t,y , g\right)\cdot\left( x,t,h\right)= \left( x , y, g\cdot h\right),$$
for all $\left(  t,y,g\right),\left(x,t,h\right) \in   X\times X \times G$. This groupoid is called \textit{trivial groupoid on $X$ with group $G$}.
\end{example}

\end{example}

\noindent{Next, let us describe the crucial example of groupoid for the purpose of this paper.}
\begin{example}\label{8}
\rm
Let $A$ be a vector bundle on a manifold $M$. Denote by $A_{z}$, the fibre of $A$ over a $z \in M$. Then, the set $\Phi \left(A\right)$, consisting of all linear isomorphisms $L_{x,y}: A_{x} \rightarrow A_{y}$ for any $x,y \in M$, may be endowed with the structure of groupoid with structure maps,
\begin{itemize}
\item[(i)] $\alpha\left(L_{x,y}\right) = x$
\item[(ii)] $\beta\left(L_{x,y}\right) = y$
\item[(iii)] $L_{y,z} \cdot G_{x,y} = L_{y,z} \circ G_{x,y}, \ L_{y,z}: A_{y} \rightarrow A_{z}, \ G_{x,y}: A_{x} \rightarrow A_{y}$
\end{itemize}
We will call this groupoid as the \textit{frame groupoid on $A$}.\\
A particular relevant case is the \textit{1-jets groupoid on} $M$ and it arises when $A$ is the tangent bundle $TM$ of $M$. This groupoid is denoted by $\Pi^{1} \left(M,M\right)$. Notice that any isomorphism $L_{x,y}: T_{x}M \rightarrow T_{y}M$ may be written as a $1-$jet $j_{x,y}^{1} \psi$ of a local diffeomorphism $\psi$ from $M$ to $M$. Notice that the $1-$jet $j_{x,y}^{1} \psi$ may be identified with the tangent map $T_{x}\psi: T_{x}M \rightarrow T_{y}M$ (see \cite{SAUND} for details).
\end{example}
\begin{definition}\label{58}
\rm
Let $\Gamma \rightrightarrows M$ be a groupoid with $\alpha$ and $\beta$ the source map and target map, respectively. For each $x \in M$, the set
$$\Gamma^{x}_{x}= \beta^{-1}\left(x\right) \cap \alpha^{-1}\left(x\right),$$
is called the \textit{isotropy group of} $\Gamma$ at $x$. The set
$$\mathcal{O}\left(x\right) = \beta\left(\alpha^{-1}\left(x\right)\right) = \alpha\left(\beta^{-1}\left( x\right)\right),$$
is called the \textit{orbit} of $x$, or \textit{the orbit} of $\Gamma$ through $x$.
\end{definition}

\noindent{Observe that the isotropy groups inherits a \textit{bona fide} group structure.}
\begin{definition}
\rm
If $\mathcal{O}\left(x\right) = M$ for all $x \in M$ (or equivalently $\left(\alpha,\beta\right) : \Gamma  \rightarrow M \times M$ is a surjective map) the groupoid $\Gamma \rightrightarrows M$ is called \textit{transitive}. The sets,
$$ \alpha^{-1} \left(x \right) = \Gamma_{x}, \ \ \ \ \ \beta^{-1} \left(x \right) = \Gamma^{x},$$
are called $\alpha-$\textit{fibre at} $x$ and $\beta-$\textit{fibre at $x$}, respectively. We will denote
$$\Gamma_{x}^{y} = \Gamma_{x} \cap \Gamma^{y},$$
for all $x,y \in M$.
\end{definition}

\begin{definition}\label{9}
\rm
Let $\Gamma \rightrightarrows M$ be a groupoid. We may define the \textit{left translation by $g \in \Gamma$} as the map $L_{g} : \Gamma^{\alpha\left(g\right)} \rightarrow \Gamma^{\beta\left(g\right)}$, given by
$$ h \mapsto  g \cdot h .$$
We may define the \textit{right translation} by $g$, $R_{g} : \Gamma_{\beta\left(g\right)} \rightarrow \Gamma_{ \alpha \left(g\right)}$, analogously. 
\end{definition}
\noindent{Note that, the identity map on $\Gamma^{x}$ may be written as the following translation map,}
\begin{equation}\label{10} 
Id_{\Gamma^{x}} = L_{\epsilon \left(x\right)}.
\end{equation}
For any $ g \in \Gamma $, the left (resp. right) translation on $g$, $L_{g}$ (resp. $R_{g}$), is a bijective map with inverse $L_{g^{-1}}$ (resp. $R_{g^{-1}}$).\\\\

\begin{definition}
\rm
A \textit{Lie groupoid} is a groupoid $\Gamma \rightrightarrows M$ such that $\Gamma$ is a smooth manifold, $M$ is a smooth manifold and the structure maps are smooth. Furthermore, the source and the target map are submersions.\\
A \textit{Lie subgroupoid} of $\Gamma \rightrightarrows M$ is a Lie groupoid $\Gamma' \rightrightarrows M'$ such that it is a subgroupoid of $\Gamma$ satisfying that $\Gamma' $ and $M'$ are submanifolds of $\Gamma$ and $M$ respectively.
\end{definition}

Notice that the following statements are immediate:
\begin{itemize}
    \item $\epsilon$ is an injective immersion.
    \item For each $g \in \Gamma$, the left translation $L_{g}$ (resp. right translation $R_{g}$) is a diffeomorphism, for all $g \in \Gamma$.
    \item For each $k \in \mathbb{N}$, $\Gamma_{\left(k\right)}$ is a smooth manifold, for all $k \in \mathbb{N}$.
    \item The $\beta-$fibres and the $\alpha-$fibres are closed submanifolds of $\Gamma$.
\end{itemize}

As firt examples, any Lie group $G$ (example \ref{5}), the pair groupoid of a manifold $M$ (example \ref{7}), and any trivial groupoid on a manifold $M$ with a Lie group $G$ (example \ref{truvialG}) are Lie groupoids.

\begin{example}\label{15}
\rm
The frame groupoid $\Phi \left( A \right)$ on a vector bundle $A$ (example \ref{8}) is a Lie groupoid . Let us consider two local coordinates, $\left(x^{i}\right)$ and $\left(y^{j}\right)$, on open neighbourhoods $U, V \subseteq M$, respectively, and two local basis of sections of $A_{U}$ and $A_{V}$, $\{\alpha_{p}\}$ and $\{ \beta_{q}\}$, respectively. The correspondent local coordinates $\left(x^{i} \circ \pi, \alpha^{p}\right)$ and $\left(y^{j} \circ \pi, \beta^{q}\right)$ on $A_{U}$ and $A_{V}$ are given by
\begin{itemize}
\item For any $a \in A_{U}$,
$$ a =  \alpha^{p}\left(a\right) \alpha_{p}\left(x^{i}\left(\pi \left(a\right)\right)\right).$$\\
\item For any $a \in A_{V}$,
$$ a =  \beta^{q}\left(a\right) \beta_{q}\left(y^{j}\left(\pi \left(a\right)\right)\right).$$
\end{itemize}
\noindent{Then, we can construct a local coordinate system on $\Phi \left(A\right)$
$$ \Phi \left(A_{U,V}\right) : \left(x^{i} , y^{j}_{i}, y^{j}_{i}\right),$$
where, $A_{U,V} = \alpha^{-1}\left(U\right) \cap \beta^{-1}\left(V\right)$ and for each $L_{x,y} \in \alpha^{-1}\left(x\right) \cap \beta^{-1}\left(y\right) \subseteq \alpha^{-1}\left(U\right) \cap \beta^{-1}\left(V\right)$, we have}
\begin{itemize}\label{16}
\item $x^{i} \left(L_{x,y}\right) = x^{i} \left(x\right)$.
\item $y^{j} \left(L_{x,y}\right) = y^{j} \left( y\right)$.
\item $y^{j}_{i}\left( L_{x,y}\right)  = A_{L_{x,y}}$, where $A_{L_{x,y}}$ is the associated matrix to the induced map of $L_{x,y}$ using the local coordinates $\left(x^{i} \circ \pi, \alpha^{p}\right)$ and $\left(y^{j} \circ \pi, \beta^{q}\right)$.
\end{itemize}
In the particular case of the $1-$jets groupoid on $M$, $\Pi^{1}\left(M,M\right)$, the local coordinates will be denoted as follows
\begin{equation}\label{17}
\Pi^{1}\left(U,V\right) : \left(x^{i} , y^{j}, y^{j}_{i}\right),
\end{equation}
where, for each $ j^{1}_{x,y} \psi \in \Pi^{1}\left(U,V\right)$
\begin{itemize}
\item $x^{i} \left(j^{1}_{x,y} \psi\right) = x^{i} \left(x\right)$.
\item $y^{j} \left(j^{1}_{x,y}\psi \right) = y^{j} \left( y\right)$.
\item $y^{j}_{i}\left( j^{1}_{x,y}\psi\right)  = \dfrac{\partial \left(y^{j}\circ \psi\right)}{\partial x^{i}_{| x} }$.
\end{itemize}

\end{example}

In the application to continuum mechanics, we will introduce the \textit{material groupoid}, which will be defined as a subgroupoid of special cases of the frame groupoid. In particular, we will deal with the $1-$jets groupoid $\Pi^{1}\left(\mathcal{B}, \mathcal{B}\right)$ on a manifold $\mathcal{B}$ (body) and a frame groupoid $\Phi \left( \mathcal{V} \right)$ of the vertical bundle $\mathcal{V}$ of a given vector bundle $\mathcal{C}$ (material evolution).\\\\

\noindent{From now on, we will deal with a (not necessarily a Lie) subgroupoid $\overline{\Gamma}  \rightrightarrows M$ of a Lie groupoid $\Gamma  \rightrightarrows M$,}\\

\begin{center}
 \begin{tikzcd}[column sep=huge,row sep=huge]
\overline{\Gamma}\arrow[r, hook, "j"] \arrow[rd, shift right=0.5ex] \arrow[rd, shift left=0.5ex]&\Gamma \arrow[d, shift right=0.5ex] \arrow[d, shift left=0.5ex] \\
& M 
 \end{tikzcd}
\end{center}

\noindent{where $j$ is the inclusion map. We will also denote by $\overline{\alpha}$, $\overline{\beta}$, $\overline{\epsilon}$ and $\overline{i}$ the restrictions of the structure maps $\alpha$, $\beta$, $\epsilon$ and $i$ of $\Gamma$ to $\overline{\Gamma}$.\\\\

\noindent{In what follows, we will construct of the so-called \textit{characteristic distribution} $A \overline{\Gamma}^{T}$ (see \cite{VMMDME, CHARDIST}).}
\begin{definition}
\rm
An \textit{admissible} (local) vector field $\Theta \in \mathfrak X_{loc} \left( \Gamma \right)$ on $\Gamma$ for the couple $\left( \Gamma , \overline{\Gamma} \right)$ is a (local) vector field on $\Gamma$ satisfying that,
\begin{itemize}
\item[(i)] $\Theta$ is tangent to the $\beta-$fibres, 
$$ \Theta \left( g \right) \in T_{g} \beta^{-1} \left( \beta \left( g \right) \right),$$
for all $g$ in the domain of $\Theta$.
\item[(ii)] $\Theta$ is invariant by left translations,
$$ \Theta \left( g \right) = T_{\epsilon \left( \alpha \left( g \right) \right) } L_{g} \left( \Theta \left( \epsilon \left( \alpha \left( g \right) \right) \right) \right),$$
for all $g $ in the domain of $\Theta$.
\item[(iii)] The (local) flow $\varphi^{\Theta}_{t}$ of $\Theta$ satisfies
$$\varphi^{\Theta}_{t} \left( \epsilon \left( x \right)\right) \in \overline{\Gamma}, $$
for all $x \in M$.
\end{itemize}
\end{definition}

In other words, an admissible vector field is a left invariant vector field on $\Gamma$ whose flow at the identities is totally contained in $\overline{\Gamma}$.\\
Moreover, a vector field $\Theta$ of $\Gamma$ satisfies conditions (i) and (ii) if, and only if, its local flow $\varphi^{\Theta}_{t}$ is left-invariant or, equivalently,
$$ L_{g} \circ \varphi^{\Theta}_{t} = \varphi^{\Theta}_{t} \circ L_{g}, \ \forall g,t.$$
Therefore, condition (iii) is equivalent to the following one,
\begin{itemize}
\item[(iii)'] The (local) flow $\varphi^{\Theta}_{t}$ of $\Theta$ at $\overline{g}$ is totally contained in $\overline{\Gamma}$, for all $\overline{g} \in \overline{\Gamma}$.
\end{itemize}
Hence, the admissible vector fields are the left-invariant vector fields on $\Gamma$ whose integral curves are confined inside or outside $\overline{\Gamma}$.\\
The family of admissible vector fields for the couple $\left( \Gamma , \overline{\Gamma} \right)$ is denoted by $\mathcal{C}_{\left( \Gamma , \overline{\Gamma} \right)}$, or simply, $\mathcal{C}$ if there is no danger of confusion.\\

\begin{definition}
\rm
The \textit{characteristic distribution of $\overline{\Gamma}$}, dentoted by $A \overline{\Gamma}^{T}$ is the smooth distribution on $\Gamma$ linearly generated by the admissible (local) vector fields.
\end{definition}
\noindent{Namely, for each $ g \in \Gamma$, the fibre at $g$ is given by 
$$A \overline{\Gamma}^{T}_{g} \ = \ Span \{ \Theta \left( g \right) \ : \ \Theta \ \text{is an admissible vector field}\}$$
Observe that, for all $g \in \Gamma$, the zero vector $0_{g} \in T_{g} \Gamma$ is contained in the fibre $A \overline{\Gamma}^{T}_{g}$ of the distribution at $g$ (we remit to \cite{CHARDIST,GENHOM} for non trivial examples).} Then, the distribution $A\overline{\Gamma}^{T}$ generated by the vector spaces $A\overline{\Gamma}^{T}_{g}$ is the \textit{characteristic distribution of $\overline{\Gamma}$}.\\
\begin{remark}
\rm
This construction of the characteristic distribution associated to a subgroupoid $\overline{\Gamma}$ of a Lie groupoid $\Gamma$ may be thought as a generalization of the construction of the associated Lie algebroid to a given Lie groupoid (see \cite{KMG}).\\
\end{remark}

The algebraic structure associated to a groupoid allows us to define more objects. Particularly, one of them is a smooth distribution over the base $M$ denoted by $A \overline{\Gamma}^{\sharp}$. The one is a ``\textit{differentiable}" correspondence $A\overline{\Gamma}$ which associates to any point $x$ of $M$ a vector subspace of $T_{\epsilon \left( x \right) } \Gamma$. Both constructions are characterized by the following diagram

\begin{large}
\begin{center}
 \begin{tikzcd}[column sep=huge,row sep=huge]
\Gamma\arrow[r, "A \overline{\Gamma}^{T}"] &\mathcal{P} \left( T \Gamma \right) \arrow[d, "T\alpha"] \\
 M \arrow[u,"\epsilon"] \arrow[r,"A \overline{\Gamma}^{\sharp}"] \arrow[ru,dashrightarrow, "A \overline{\Gamma}"]&\mathcal{P} \left( T M \right)
 \end{tikzcd}
\end{center}
\end{large}

%

\vspace{15pt}
\noindent{where $\mathcal{P} \left( E \right)$ defines the power set of $E$. Therefore, for any $x \in M$, the fibres are given by,}
\begin{eqnarray*}
A \overline{\Gamma}_{x} &=&  A \overline{\Gamma}^{T}_{\epsilon \left( x \right)}\\
A \overline{\Gamma}^{\sharp}_{x}  &=& T_{\epsilon \left( x \right) } \alpha \left( A \overline{\Gamma}_{x} \right)
\end{eqnarray*}
\noindent{The distribution $A \overline{\Gamma}^{\sharp}$ is called \textit{base-characteristic distribution of $\overline{\Gamma}$}. Observe that, those distributions are possibly singular.}\\
Notice that, taking into account that $A \overline{\Gamma}^{T}$ is locally generated by left-invariant vector field, we have that for each $g \in \Gamma$,
$$ A \overline{\Gamma}^{T}_{g} = T_{\epsilon \left( \alpha \left( g \right) \right)} L_{g} \left( A \overline{\Gamma}^{T}_{\epsilon \left( \alpha \left( g \right)\right)} \right),$$
i.e., the characteristic distribution is \textit{left-invariant}.}\\
\begin{theorem}[\cite{VMMDME,CHARDIST}]\label{10.24}
Let $\Gamma \rightrightarrows M$ be a Lie groupoid and $\overline{\Gamma}$ be a subgroupoid of $\Gamma$ (not necessarily a Lie groupoid) over $M$. Then, the characteristic distribution $A \overline{\Gamma}^{T}$ is integrable and its associated foliation $\overline{\mathcal{F}}$ of $\Gamma$ satisfies that $\overline{\Gamma}$ is a union of leaves of $\overline{\mathcal{F}}$.
\end{theorem}
This result is a consequence of the celebrated Stefan-Sussman's theorem \cite{PS,HJS} which deals with the integrability of singular distributions. Each leaf at a point $g \in \Gamma$ will be denoted by $\overline{\mathcal{F}} \left( g \right)$ and the \textit{characteristic foliation of $\overline{\Gamma}$} will be given by the family of the leaves of $\overline{\mathcal{F}}$ at points of $\overline{\Gamma}$. The foliation $\overline{\mathcal{F}}$ satisfies that
\begin{itemize}
\item[(i)] For any $g \in \Gamma$,
$$\overline{\mathcal{F}} \left( g \right) \subseteq \Gamma^{\beta \left( g \right)}.$$
Indeed, if $g \in \overline{\Gamma}$, then
$$\overline{\mathcal{F}} \left( g \right) \subseteq \overline{\Gamma}^{\beta \left( g \right)}.$$
\item[(ii)] \textit{Left-invariance:} for any $g ,h \in \Gamma$ such that $\alpha \left( g \right) = \beta \left( h \right)$, we have
$$\overline{\mathcal{F}} \left( g \cdot h\right) = g \cdot \overline{\mathcal{F}} \left(  h\right).$$
\end{itemize}
It is important to point out that the leaves of the characteristic foliation covers $\overline{\Gamma}$ but, however, it is not exactly a foliation of $\overline{\Gamma}$ (because $\overline{\Gamma}$ is not necessarily a manifold).\\ 
As a summary, without any assumption of differentiability over $\overline{\Gamma}$, we have that $\overline{\Gamma}$ is union of leaves of a foliation of $\Gamma$. This provides some kind of ``\textit{differentiable}" structure over $\overline{\Gamma}$.\\

\noindent{Notice that, analogously to theorem \ref{10.24}, we may prove that the base-characteristic distribution $A \overline{\Gamma}^{\sharp}$ is integrable. Thus, we will denote the foliation which integrates the base-characteristic distribution over the base $M$ by $\mathcal{F}$. For each point $x \in M$, the leaf of $\mathcal{F}$ containing $x$ will be denoted by $\mathcal{F} \left( x \right)$. $\mathcal{F}$ will be called the \textit{base-characteristic foliation of $\overline{\Gamma}$}.}\\
\begin{example}\label{10.41}
\rm
Let $\sim$ be an equivalence relation on a manifold $M$, i.e., a binary relation that is reflexive, symmetric and transitive. Then, define the subset $\mathcal{O}$ of $M \times M$ given by
\begin{equation}
\mathcal{O}:= \{ \left( x,y \right) \  : \    x \sim y \}.
\end{equation}
Hence, $\mathcal{O}$ is a subgroupoid of $M \times M$ over $M$. In fact, this is equivalent to the properties reflexive, symmetric and transitive. For each $x \in M$, we denote by $\mathcal{O}_{x}$ to the orbit around $x$,
$$\mathcal{O}_{x}:= \{ y  \  : \    x \sim y \}.$$
Notice that the orbits divide $M$ into a disjoint union of subsets. However, these are not (necessarily) submanifolds.\\
On the other hand, the base-characteristic foliation gives us a foliation $\mathcal{F}$ of $M$ such that
$$ \mathcal{F} \left( x \right) \subseteq \mathcal{O}_{x}, \ \forall x \in M.$$
So, consider any arbitrary equivalence relation on a manifold $M$. Maybe the orbits are not manifolds but we have proved that \textit{we may divide $M$ in a maximal foliation such that any orbit is a union of leaves.} This foliation is maximal in the sense that there is no any other coarser foliation of $M$ whose leaves are contained in the orbits (see theorem \ref{10.20} and corollary \ref{10.39}).
\end{example}

\noindent{Next, we will show that the leaves of $\mathcal{F}$ may be endowed with even more geometric structure. Indeed, we will construct a Lie groupoid structure over each leaf of $\mathcal{F}$.}\\
For each $x \in M$, let us consider the minimal transitive groupoid $\overline{\Gamma} \left( \mathcal{F} \left( x \right) \right)$ generated by the elements of $\overline{\mathcal{F}} \left( \epsilon \left( x \right) \right) $. In fact, we may prove that
\begin{equation}\label{10.17}
 \overline{\Gamma} \left( \mathcal{F} \left( x \right) \right) = \sqcup_{\overline{g} \in \overline{\mathcal{F}} \left( \epsilon \left( x \right) \right)} \overline{\mathcal{F}} \left( \epsilon \left( \alpha \left( \overline{g} \right) \right) \right),
\end{equation}
i.e., $\overline{\Gamma} \left( \mathcal{F} \left( x \right) \right)$ can be described as a disjoint union of fibres at the identities.
\begin{theorem}[\cite{VMMDME,CHARDIST}]\label{10.20}
For each $x \in M$ there exists a transitive Lie subgroupoid $\overline{\Gamma} \left( \mathcal{F} \left( x \right) \right)$ of $\Gamma$ with base $\mathcal{F} \left( x \right)$.

\end{theorem}
Thus, we have divided the manifold $M$ into leaves $\mathcal{F} \left( x \right)$ which have a maximal structure of transitive Lie subgroupoids of $\Gamma$. The following result provides us an intuition about the maximality condition which satisfies the characteristic foliation and the base-characteristic foliation.\\

\begin{corollary}[\cite{VMMDME}]\label{10.39}
Let $\mathcal{H}$ be a foliation of $M$ such that for each $x \in M$ there exists a transitive Lie subgroupoid $\Gamma \left( x \right)$ of $\Gamma$ over the leaf $\mathcal{H} \left( x \right)$ contained in $\overline{\Gamma}$ whose family of leaves defines a foliation on $\Gamma$. Then, the base-characteristic foliation $\mathcal{F}$ is coarser than $\mathcal{H}$, i.e.,
$$ \mathcal{H} \left( x \right) \subseteq \mathcal{F} \left( x \right) , \ \forall x \in M.$$
Futhermore, it satisfies that
$$ \Gamma \left( x \right) \subseteq \overline{\Gamma} \left( \mathcal{F} \left( x \right) \right).$$
\end{corollary}
As a consequence we have that $\overline{\Gamma}$ is a transitive Lie subgroupoid of $\Gamma$ if, and only if, $M = \mathcal{F} \left( x \right)$ and $\overline{\Gamma} =\overline{\Gamma} \left( \mathcal{F} \left( x \right) \right)$ for some $x \in M$.\\

Thus, summarizing, for a fixed subgroupoid $\overline{\Gamma}$ of a Lie groupoid $\Gamma$ we have available two canonical foliations, $\overline{\mathcal{F}}$ and $\mathcal{F}$ which endow to $\Gamma$ of some kind of maximal differentiable structure. To study more properties of the characteristic distribution, we recommend \cite{VMMDME}.\\
Apart from example \ref{10.41}, we may study several relevant applications of the characteristic distribution. In \cite{VMMDME} we may find some of them. Here we are mainly interested in one of them, the so-called \textit{material distributions}, which will be presented in what follows.

\section{Material evolution of simple bodies}\label{partelasticmat}

We will now present the notion of \textit{simple material} mainly the references \cite{CTRUE,CCWAN}.

\begin{definition}\label{45body}
\rm
A \textit{(deformable) body} is given by an oriented manifold $\mathcal{B}$ of dimension $3$ which can be covered by just one chart. The points of $\mathcal{B}$ will be called \textit{body points} or \textit{material particles} and will be denoted by using capital letters ($X,Y,Z \in \mathcal{B}$). A \textit{sub-body} of $\mathcal{B}$ is an open subset $\mathcal{U}$ of the manifold $\mathcal{B}$.
\end{definition}
A \textit{configuration} of the body $\mathcal{B}$ is an embedding $\phi : \mathcal{B} \rightarrow \mathbb{R}^{3}$ and an \textit{infinitesimal configuration at a particle $X$} is given by the $1-$jet $j_{X,\phi \left(X\right)}^{1} \phi$ where $\phi$ is a configuration of $\mathcal{B}$. The image $\phi \left( \mathcal{B}\right)$ is called the \textit{region occupied by the body $\mathcal{B}$ in the configuration} $\phi$. The points on the Euclidean space $\mathbb{R}^{3}$ will be called \textit{spatial points} and will be denoted by lower case letters ($x,y,z\in \mathbb{R}^{3}$).\\
We will now fix a configuration, denoted by $\phi_{0}$, called \textit{reference configuration}. The image will be denoted by $\mathcal{B}_{0} = \phi_{0} \left( \mathcal{B} \right)$. Coordinates in the reference configuration will be denoted by $X^{I}$, while any other coordinates will be denoted by $x^{i}$.\\
Notice that, in \cite{CTRUE} a body is defined simply as a three-dimensional manifold. Nevertheless, without loss of generality, we will adopt the above definition of body \ref{45body} which is used in \cite{CCWAN}.  On the other hand, in \cite{JEMARS} a body is simple defined as an open subset of $\mathbb{R}^{3}$, so the body is identified with its image in $\mathbb{R}^{3}$ via a reference configuration.\\
Any change of configurations $\kappa = \phi_{1} \circ \phi_{0}^{-1}$ or, equivalently a diffeomorphism from $\mathcal{B}_{0}$ to any other open subset $\mathcal{B}_{1}$ of $\mathbb{R}^{3}$, is called a \textit{deformation}. Analogously, an \textit{infinitesimal deformation at $\phi_{0}\left(X\right)$} is given by a $1-$jet $j_{\phi_{0}\left(X\right) , \phi \left(X\right)}^{1} \kappa$, where $\kappa$ is a deformation.\\
The change of the body in time will be modelized by the \textit{body-time manifold}, which is defined as the fibre bundle $ \mathcal{C} = \mathbb{R} \times \mathcal{B}$ over $\mathbb{R}$. Then, a \textit{history} is given by a fibre bundle embedding $\Phi : \mathcal{C} \rightarrow \mathbb{R} \times \mathbb{R}^{3}$ over the identity.\\
Notice that $\Phi$ can be seen as a differentiable family of configurations $\phi_{t} : \mathcal{B} \rightarrow \mathbb{R}^{3}$ such that
\begin{equation}\label{1.3.4.5}
\phi_{t} \left( x \right) = pr_{\mathbb{R}^{3}}\circ \Phi \left( t , x \right), \ \forall t \in \mathbb{R}, \ \forall x \in \mathcal{B},
\end{equation}
where $pr_{\mathbb{R}^{3}}: \mathbb{R}\times \mathbb{R}^{3} \rightarrow \mathbb{R}^{3}$ is the projection on $\mathbb{R}^{3}$. Thus, $\Phi$ represent the evolution of the body in time $t$ in such a way that the configuration of $\mathcal{B}$ at time $t$ is $\phi_{t}$. Then, at each instant of time $t$, one may consider the infinitesimal configuration at time $t$, $1-$jet $j_{X, \phi_{t}\left(X \right)}^{1} \phi_{t}$.\\

\noindent{Of course, the instrinsic properties of the body will play an important role in continuum mechanics. One of the most characteristic contributions of the work of W. Noll \cite{WNOLLTHE}, was the introduction of the \textit{mechanical response}; a differentiable map characterizing the internal properties of the material. In the case of \textit{elastic simple bodies}, or simply \textit{simple bodies}, \cite{CCWAN} we will assume that the mechanical response depends on a particle only on the \textit{infinitesimal deformation} at the same particle. Then, following \cite{CTRUE, EPSTEIN201572,EPSBOOK2}, we will assume that, for a fixed reference configuration $\phi_{0}$, the constitutive response at each material particle $X$ and at each instant of time $t$, the mechanical response may be characterized by one (or more) functions depending on the associated matrices $F$ to the infinitesimal configurations $j_{X, \phi_{t}\left(X \right)}^{1} \phi_{t}$ at particle $X$ and time $t$.\\
\begin{definition}
\rm
Let $\mathcal{B}$ be a simple body with $ \mathcal{C} = \mathbb{R} \times \mathcal{B}$ as the associated body time manifold. Then, for a fixed reference configuration $\phi_{0}$, the \textit{mechanical response} will be a differentiable map,
$$ W :  \mathcal{C} \times Gl\left(3 , \mathbb{R}\right)  \rightarrow V,$$
where $V$ is again a real vector space.
\end{definition}
Generally, in continuum mechanics, the contact forces at a particle $X$ at an instant $t$ in a given configuration $\phi$ are characterized by a symmetric second-order tensor $T_{t,X,\phi}$ on $\mathbb{R}^{3}$, which is called the stress tensor. Then, the mechanical response is given by the following equation:
$$W \left( t, X , F \right) = T_{t, X,\phi},$$
where $F$ is the $1-$jet at $\phi_{0} \left( X \right)$ of $\phi \circ \phi_{0}^{-1}$. Namely, in general, $V$ will be the space of \textit{stress tensors} \cite{STRESS} although, in this paper, we will only be interested in the structure of vector space of $V$.
Notice that, the definition of the mechanical response permits us to compare material responses at different particles at different instants of time. Relevant examples are given by the volumetric growth and remodeling of biological tissues, such as bone and muscle \cite{RODRIGUEZ1994455}.}\\

Observe that, the construction of the mechanical response seems to be constrained to the fixed reference configuration. To clarify this dependence we have the \textit{rule of change of reference configuration}.\\
Consider a different configuration $\phi_{1}$ and $W_{1}$ its associated mechanical response. Then, it will be imposed that
\begin{equation}\label{1.5.2}
 W_{1} \left(t, X , F \right) = W \left( t, X , F \cdot C_{01} \right),
\end{equation}
for all regular matrix $F$ where $C_{01}$ is the associated matrix to the $1-$jet at $\phi_{0} \left( X \right)$ of $\phi_{1} \circ \phi_{0}^{-1}$. Equivalently,
\begin{equation}\label{1.4.2}
 W \left( t,X , F_{0} \right) = W_{1} \left(t, X , F_{1} \right),
\end{equation}
where $F_{i}$, $i=0,1$, is the associated matrix to the $1-$jet at $\phi_{i} \left( X \right)$ of $\phi \circ \phi_{i}^{-1}$ with $\phi$ a configuration. Therefore, Eq. (\ref{1.5.2}) permits us to define $W$ over the space of (local) histories which is \textit{independent on the chosen reference configuration}. In particular, for all history $\Phi = \phi_{t}$, we will define
\begin{equation}\label{1.4.5.5}
    W \left( t , X , \Phi \right) = W \left(t,  j^{1}_{X,x} \phi_{t} \right) = W \left( t, X , F_{t} \right),
\end{equation}
where $F_{t}$ is the associated matrix to the $1-$jet $j^{1}_{\phi_{0} \left( X \right),x}\left( \phi_{t} \circ \phi_{0}^{-1}\right)$ at $\phi_{0} \left( X \right)$.\\
Observe that, for all $t$ the manifold $\{t\}\times \mathcal{B}$ inherits the structure of simple body by restricting the mechanical response $W$ to the history of deformations at the same instant $t$ (see \cite{VMMDME}), i.e.,
$$ W_{t} : \{t\}\times \mathcal{B} \times Gl\left(3 , \mathbb{R}\right)  \rightarrow V.$$
This body will be called \textit{state $t$ of the body $\mathcal{B}$}. Thus, we may think about $W$ as a differentiable curve of mechanical responses, each one of these over the corresponding state of the body. As long as it invites no confusion, we will refer to the simple body $\{0\} \times \mathcal{B}$ as the material body $\mathcal{B}$.\\
On the other hand, it is also important to say that the mechanical response defines a structure of material evolution on any sub-body $\mathcal{U}$ of the body $\mathcal{B}$ by restriction. Nevertheless, we will need to relax the definition of ``\textit{material evolution}'' to permit variation of material submanifolds along time.

\begin{definition}\label{materialsubevol56}
\rm
An \textit{evolution material} for a \textit{submanifold} (or \textit{body-time generalized sub-body}) of $\mathcal{C}$ is a submanifold $\mathcal{M}$ of $\mathcal{C}$.
\end{definition}
Thus, let us consider an instant $t$. Then, we have that the \textit{state $t$ of the material submanifold} is 
$$\left(\{t\}\times \mathcal{B}  \right) \cap \mathcal{M} = \{t\}\times \mathcal{M}_{t},$$
for a submanifold $\mathcal{M}_{t}$ of $\mathcal{B}$. Hence, varying $t$, the model permits variations in the ``\textit{shape}'' of $\mathcal{M}_{t}$.\\
Maybe the cornerstone of the thesis of W. Noll \cite{WNOLLTHE} is the use of the so-called \textit{material isomorphisms}. This notion arises from the need to respond the following question: when are two material points made of the same material? Obviously, we could say that two points are made of the same material if the constitutive response is exactly the same at both points. However, this is not enough general to answer the question.
\begin{definition}\label{1.33.2}
\rm
Let $\mathcal{C}$ be a body-time manifold. Two pairs $\left( t, X \right), \left( s ,Y\right) \in \mathcal{C}$ are said to be \textit{materially isomorphic} if there exists a local diffeomorphism $\psi$ from an open neighbourhood $\mathcal{U} \subseteq \mathcal{B}$ of $X$ to an open neighbourhood $\mathcal{V} \subseteq \mathcal{B}$ of $Y$ such that $\psi \left(X\right) =Y$ and
\begin{equation}\label{1.3.2}
W \left( t, X , F \cdot P \right) = W \left( s, Y , F \right),
\end{equation}
for all infinitesimal deformation $F$ where $P$ is given by the Jacobian matrix of $\phi_{0} \circ \psi \circ \phi_{0}^{-1}$ at $\phi_{0} \left( X \right)$. The triples given by $\left(t,s, j_{X,Y}^{1}\psi \right)$, with $\psi$ satisfying Eq. (\ref{1.3.2}) are called \textit{time-material isomorphisms} (or \textit{material isomorphisms} if there is no danger of confusion) from $\left(t,X\right)$ to $\left(s,Y\right)$. A material isomorphism from $\left(t,X\right)$ to itself is called a \textit{time-material symmetry} or \textit{material symmetry}.
\end{definition}
\noindent{Let us denote by $G \left(t, X \right)$ to the group of all material symmetries at $\left(t,X\right)$.} Thus, the notion of being ``\textit{material isomorphic}'' is the mathematical formulation of the intuitive idea of being made of the same of the material, in such a way that two material points (at maybe different instants) are made of the same material if, and only if, they are materially isomorphic.
\begin{proposition}
Let $\mathcal{C}$ be a body-time manifold. Two body pairs $\left(t,X\right)$ and $\left(s,Y\right)$ are materially isomorphic if, and only if, there exist two (local) configurations $\phi_{1}$ and $\phi_{2}$ such that
\begin{equation}\label{isom54}
W_{1} \left( t,X , F \right)= W_{2} \left( s,Y , F \right), \ \forall F,
\end{equation}
where $W_{i}$ is the mechanical response associated to $\phi_{i}$ for $i=1,2$.
\begin{proof}

Consider $j_{X,Y}^{1}\psi$ a material isomorphism from $\left(t,X\right)$ to $\left(s,Y\right)$. We only have to choose $\phi_{1} = \phi_{0}$ and
$$\phi_{2} = \phi_{1}\circ \psi$$
\end{proof}
\end{proposition}
So, this result shows the reason why the mathematical notion of material isomorphism answers to the question posed at the beginning of the section \ref{partelasticmat}. In particular, two points are materially isomorphic if their constitutive properties are equal (up to the choice of the reference configuration). Roughly speaking, we may turns neighbourhoods of a material particle $X$ at an instant $t$ into neighbourhoods of the material particle $Y$ at the other instant $s$, such that the \textit{stress required to effect any given deformation} of those neighborhoods is the same for each.

\begin{remark}
\rm
It is important to note that, in \cite{CTRUE}, authors define the notion of \textit{materially isomorphic} by using equation (\ref{isom54}), next to an added condition refered to the \textit{mass density}.\\
A non-negative scalar measure, $\mathfrak{m}$ defined on the body manifold $\mathcal{B}$ is called the \textit{mass distribution} of the body. Then, $\mathfrak{m}$ induces a measure over each image $\phi \left( \mathcal{B}\right)$ of $\mathcal{B}$ via a configuration $\phi$ denoted by $\mathfrak{m}_{\phi}$ which is assumed to be absolutely continuous with respect to the Lebesgue measure in Euclidean space $\mathbb{R}^{3}$. Therefore, by using the Radon-Nikodym theorem, we may construct a density $\rho_{\phi}$, called \textit{mass density associated to} $\phi$. In fact, the relation between the mass and the density is the following,

$$\mathfrak{m} \left( \mathcal{P} \right) = \int_{\phi \left( \mathcal{P}\right)} \rho_{\phi} \  d v,$$
for each every measurable subset $\mathcal{P}$ of $\mathcal{B}$. Here, the integral is defined in terms of Lebesgue measure in Euclidean space.\\
Finally, let us consider a suitable measurable subsets $\mathcal{P}_{k}$ having only the particle $X$ in common and satisfying that
$$\lim_{k \to \infty} \mathfrak{v}\left( \mathcal{P}_{k} \right) = 0,$$
where $\mathfrak{v}$ defines the \textit{volume map} respect to Lebesgue measure. Then, we define
$$\rho_{\phi} \left( X \right) = \lim_{k \to \infty} \frac{\mathfrak{m}_{\phi}\left( \mathcal{P}_{k} \right)}{\mathfrak{v}\left( \mathcal{P}_{k} \right)}$$
Thus, in \cite{CTRUE}, it is imposed that two material particles $X$ and $Y$ are materially isomorphic if, and only if, it satisfies Eq. (\ref{isom54}) and
$$\rho_{\phi_{1}} = \rho_{\phi_{2}} = \text{constant}.$$
Namely, roughly speaking, material isomorphisms preserves the stress and the density. However, for the notion of material isomorphism, we will follows the philosophy of the book \cite{CCWAN}, where the authors state that, in static problems, since the 
inertia vanishes, this condition serves no purpose.

\end{remark}

\section{Evolution material geometry}\label{section4}

Let $\mathcal{V}$ be the vertical subbundle associated to the body-time manifold $\mathcal{C}$. Namely, $\mathcal{V}$ is the subbundle of the tangent bundle $T \mathcal{C}$ whose fibres are given by
$$ \mathcal{V}_{\left( t, X \right)} = \{0\} \times T_{X}\mathcal{B},$$
for all $\left( t, X \right) \in \mathcal{C}$. Then, we may consider the associated frame groupoid $\Phi \left(\mathcal{V}\right) \rightrightarrows \mathcal{C}$ introduced in example \ref{8}. In fact, the elements of $\Phi \left(\mathcal{V}\right)$ are the linear isomorphisms between fibres of $\mathcal{V}$. We may represents the elements of $\Phi \left(\mathcal{V}\right)$ in two different ways:
\begin{itemize}
    \item[1.-] Let $\Phi: \mathcal{C} \rightarrow \mathcal{C}$ be a (local) embedding of fibre bundles. Then, the triple $\left( \left( t, X \right) , \Phi \left( t , X \right) ,  j_{X, \phi_{t}\left(X \right)}^{1} \phi_{t} \right)$, with $\phi_{t}: \mathcal{B} \rightarrow \mathcal{B}$ the associated family of configurations given in Eq. (\ref{1.3.4.5}), characterizes a linear isomorphism
$$L_{\left( t,X \right), \Phi \left( t , X \right)} : \mathcal{V}_{ \left( t , X \right)} \rightarrow \mathcal{V}_{\Phi \left( t , X \right)}$$
Reciprocally, any linear isomorphism between fibres of $\mathcal{V}$ may be represented as a triple generated by a (local) embedding of fibre bundles $\Phi: \mathcal{C} \rightarrow \mathcal{C}$.

\item[2.-] Another, less intuitive but easier, way to represent an element of $\Phi \left(\mathcal{V}\right)$ is a triple $\left(t,s , j_{X,Y}^{1} \phi\right)$ with $s,t \in \mathbb{R}$, $X \in \mathcal{B}$ and $\phi$ a local diffeomorphism from $\mathcal{B}$ to $\mathcal{B}$ such that $\phi \left(X\right)= Y$.
\end{itemize}

{\noindent{To define the structure of differentiable manifold on $\Phi \left( \mathcal{V}\right)$, we will use the second representation of the elements of $\Phi \left( \mathcal{V}\right)$. So, the local coordinates of $\Phi \left( \mathcal{V}\right)$ (see example (\ref{15})) are given by}}

\begin{equation}\label{17.second2313}
\Phi \left(  \mathcal{V}_{\mathcal{U}}\right) : \left(t,s,x^{i} , y^{j}, y^{j}_{i}\right),
\end{equation}
where, for each $ \left( t , s , j_{X,Y}^{1}\phi \right) \in \Phi \left(  \mathcal{V}_{\mathcal{U}}\right) $
\begin{itemize}
\item $t \left( t , s , j_{X,Y}^{1}\phi \right) = t$.
\item $s\left( t , s , j_{X,Y}^{1}\phi \right) = s.$
\item $x^{i} \left( t , s , j_{X,Y}^{1}\phi \right) = x^{i} \left(X\right)$.
\item $y^{j} \left( t , s , j_{X,Y}^{1}\phi \right) = y^{j} \left( Y\right)$.
\item $y^{j}_{i}\left( t , s , j_{X,Y}^{1}\phi \right)  = \dfrac{\partial \left(y^{j}\circ \phi\right)}{\partial x^{i}_{| X} }$.
\end{itemize}
being $\left(x^{i}\right)$ and $\left( y^{i}\right)$ local charts defined on the open subsets of $\mathcal{B}$, $\mathcal{U}$ and $\mathcal{W}$, respectively, and $\Phi \left(  \mathcal{V}_{\mathcal{U}, \mathcal{W}} \right)$ is given by the triples $\left( t , s , j_{X,Y}^{1}\phi \right)$ such that $X \in \mathcal{U}$ and $Y \in \mathcal{W}$.\\\\

\noindent{The importance of $\Phi \left(  \mathcal{V} \right)$ lies in the fact that all the material isomorphisms (see definition \ref{1.3.2}) are elements of $\Phi \left(  \mathcal{V} \right)$. On the other hand, the mechanical response may be defined on this groupoid. Indeed, by using Eq. (\ref{1.4.5.5}), we may define $W$ on the frame groupoid of $\mathcal{V}$,}
$$W : \Phi \left(\mathcal{V}\right) \rightarrow V,$$
as follows,
$$ W \left( t , s , j_{X,Y}^{1} \phi\right) = W\left( t , X , \Phi \right),$$
such that
$$ \Phi \left( s , Y \right) = \left( s , \phi_{0} \circ \phi \left( Y \right) \right), \ \forall \left( s , Y \right) \in \mathcal{C},$$
where $\phi_{0}$ is the reference configuration. Notice that, in consequence, $W$ does not depend on the final point, i.e., for all $\left(t, X\right) ,\left( s, Y \right),\left(r , Z \right)\in \mathcal{C}$
\begin{equation}\label{92.second}
 W \left(t,s , j_{X,Y}^{1} \phi\right)  =  W \left(t,r , j_{X,Z}^{1} \left(  \phi_{0}^{-1}\circ \tau_{Z-Y} \circ \phi_{0} \circ \phi\right)\right), 
\end{equation}
for all $\left(t,s , j_{X,Y}^{1} \phi\right) \in \Phi \left(\mathcal{V}\right)$ where $\tau_{v}$ is the translation map on $\mathbb{R}^{3}$ by the vector $v$. This point of view will be useful for our purpose.\\\\
{\noindent{

On the other hand, we may define the \textit{material groupoid} of a body-time manifold with mechanical response $W$ as the largest subgroupoid $\Omega \left( \mathcal{C} \right)\rightrightarrows \mathcal{C}$ of $\Phi \left(\mathcal{V}\right)$ such that leaves $W$ invariant. More explicitly, an element of $\Phi \left(\mathcal{V}\right)$  $\left(t,s , j_{X,Y}^{1} \phi\right)$ is in the material groupoid, if and only, if}}
$$ W \left(t,r , j^{1}_{X,Z}\left( \psi \cdot \phi \right) \right) = W \left(s,r , j^{1}_{Y,Z}\psi \right) , $$
for all $\left(s,r , j^{1}_{Y,Z}\psi \right) \in \Phi \left(\mathcal{V}\right)$. In other words, $\Omega \left( \mathcal{C} \right)$ is the space of all (time-)material isomorphisms (see definition \ref{1.33.2}). This groupoid was first presented in \cite{MEPMDLSEG}.\\
The isotropy group at each $\left( t, X \right) \in \mathcal{C}$ will be denoted by $G \left( t , X \right)$ and its elements are the material symmetries at $\left( t,X \right)$. Observe that, as in the spatial case \cite{VMMDME,CHARDIST}, the resulting groupoid does not have to be a Lie subgroupoid.\\\\
\noindent{We will also define the \textit{$\left(X,Y\right)-$material groupoid} $\Omega_{X,Y} \left( \mathbb{R} \right)$ as the set of all material isomorphisms from the particle $X$ to the particle $Y$ varying the time variable. Observe that, when $X=Y$, the $\left(X,X\right)-$material groupoid $\Omega_{X,X} \left( \mathbb{R} \right)$ is a subgroupoid of the material groupoid $\Omega \left( \mathcal{C} \right)$. For each material point $X$, $\Omega_{X,X} \left( \mathbb{R} \right)$ is called \textit{$X-$material groupoid} and denoted by $\Omega_{X} \left( \mathbb{R} \right)$.\\
On the other hand, $\Omega_{X} \left( \mathbb{R} \right)$ may be consider as a subgroupoid of $\left(\mathbb{R}  \times \mathbb{R}\right) \times \Pi^{1} \left( \mathcal{B} , \mathcal{B}\right)_{X}^{X}$ on $\mathbb{R}$, where we are identifying $\mathbb{R}$ with $\mathbb{R} \times \{X\}$. Furthermore, the structure of Lie groupoid of $\left(\mathbb{R}  \times \mathbb{R}\right) \times \Pi^{1} \left( \mathcal{B} , \mathcal{B}\right)_{X}^{X}$ is given by
$$\left( s,t,j_{X,X}^{1}\phi\right) \cdot \left( r,s,j_{X,X}^{1}\psi\right) = \left( r,t,j_{X,X}^{1}\left(\phi \circ \psi \right)\right), $$
for all $\left( s,t,j_{X,X}^{1}\phi\right) , \left( r,s,j_{X,X}^{1}\psi\right) \in  \mathbb{R}\times \mathbb{R} \times\Pi^{1} \left( \mathcal{B} , \mathcal{B}\right)_{X}^{X}$ (see example \ref{truvialG}). Again, we will use both interpretations of $\Omega_{X} \left( \mathbb{R} \right)$ along the paper.}

\begin{proposition}\label{auxiliarprop34re}
Let $\Omega \left( \mathcal{C} \right)$ be the material groupoid. If $\Omega \left( \mathcal{C} \right)$ is a Lie subgroupoid of $\Phi \left( \mathcal{V} \right)$, then for all material point $X$ we have that $\Omega_{X} \left( \mathbb{R} \right)$ is a Lie subgroupoid of $\Phi \left( \mathcal{V} \right)$.
\begin{proof}
Assume that $\Omega \left( \mathcal{C} \right)$ is a Lie subgroupoid of $\Phi \left( \mathcal{V} \right)$. Let us consider the following submersions
$$\pi_{1}: \Omega \left( \mathcal{C} \right) \rightarrow \mathcal{B} \times \mathcal{B},$$
given by
$$ \pi_{1}\left( t, s , j_{X,Y}^{1}\phi \right) = \left( X , Y \right),$$
for all $\left(t, s , j_{X,Y}^{1}\phi \right) \in \Omega \left( \mathcal{C} \right)$. Then
$$ \Omega_{X} \left( \mathbb{R} \right) = \pi_{1}^{-1}\left( X, X \right).$$

\end{proof}
\end{proposition}
So, the imposition of \textit{``being a lie groupoid''} is stronger over the material groupoid than over the $X-$material groupoids. Notice the material groupoid $\Omega \left( \mathcal{C} \right)$ encompasses the whole evolution of the body $\mathcal{B}$ and the $X-$material groupoid $\Omega_{X} \left( \mathbb{R} \right)$ codifies the evolution of the particle $X$.\\
Therefore, as a summary, we have some canonical subgroupoids ($\Omega \left( \mathcal{C} \right)$ and $\Omega_{X} \left( \mathbb{R} \right)$) of a particular Lie groupoid ($\Phi \left( \mathcal{V} \right)$), i.e., we are facing a situation which fits in the framework of the construction of the characteristic distribution (see section \ref{chardist345}).

\vspace{0.7cm}

In \cite{EvoEqu,CharFol}, the authors provides a specific representation of the correspondent characteristic distributions. In fact, the associated characteristic distribution $A \Omega\left( \mathcal{C} \right)^{T}$ to the material groupoid, which will be called \textit{material distribution} of the body-time manifold $\mathcal{C}$ is generated by the (left-invariant) vector fields $\Theta$ on $\Phi \left( \mathcal{V} \right)$ which are in the kernel of $TW$, i.e.,
\begin{equation}\label{matgrou2}
    T W \left( \Theta \right) = 0.
\end{equation}
Namely, let $\Theta$ be a left-invariant vector field on $\Phi \left( \mathcal{V} \right)$. Then,
\begin{equation}
    \Theta  \left(t,s,x^{i} , y^{j}, y^{i}_{j}\right)   = \lambda \dfrac{\partial}{\partial  t}   +   \Theta^{i}\dfrac{\partial}{\partial  x^{i} } + y^{i}_{l}\Theta^{l}_{j}\dfrac{\partial}{\partial y^{i}_{j} }
\end{equation}
respect to a local system of coordinates $\left(t,s,x^{i} , y^{j}, y^{i}_{j}\right)$ on $ \Phi \left(  \mathcal{V}_{\mathcal{U}, \mathcal{V}}\right)$ with $\mathcal{U}$ and $\mathcal{V}$ two open subsets of $\mathcal{B}$ and $ \Phi \left(\mathcal{V}_{\mathcal{U}, \mathcal{V}} \right)$ is given by the triples $\left( t, s , j_{X,Y}^{1}\phi    \right)$ in $\Phi \left( \mathcal{V} \right)$ such that $X \in \mathcal{U}$ and $Y \in \mathcal{V}$. Then, $\Theta$ is an admissible vector field for the couple $\left( \Phi \left( \mathcal{V} \right) ,\Omega \left( \mathcal{C} \right)\right)$ if, and only if,the following equations hold,
\begin{equation}\label{Eqmaterialgroupoid12}
    \lambda \dfrac{\partial W}{\partial  t}  +  \Theta^{i}\dfrac{\partial W}{\partial  x^{i} } + y^{i}_{l}\Theta^{l}_{j}\dfrac{\partial W}{\partial  y^{i}_{j} } = 0.
\end{equation}
Notice that, here $\lambda$, $\Theta^{i}$ and $\Theta^{i}_{j}$ are function depending on $t$ and $X$. Eq. (\ref{Eqmaterialgroupoid12}) is the so-called \textit{evolution equation}, which is a tool to characterize \textit{remodeling} and \textit{aging} \cite{CharFol}.\\
Thus, to construct the material distribution, we have to solve the evolution equation (\ref{Eqmaterialgroupoid12}). The base-characteristic distribution $A \Omega \left( \mathcal{C} \right)^{\sharp}$ will be called \textit{body-material distribution}.\\
The foliations associated to the material distribution and the body-material distribution is called \textit{material foliation} and \textit{body-material foliation} and they will be denoted by $\overline{\mathcal{F}}$ and $\mathcal{F}$, respectively.\\\\

\noindent{The characteristic distribution associated to the $X-$material groupoid $A \Omega_{X} \left(\mathbb{R} \right)^{T}$ is called \textit{$X-$material distribution}. Analogously, $A \Omega_{X} \left(\mathbb{R} \right)^{T}$ is generated by the (left-invariant) vector fields on $\left(\mathbb{R}  \times \mathbb{R}\right) \times \Pi^{1} \left( \mathcal{B} , \mathcal{B}\right)_{X}^{X}$ which are in the kernel of $TW_{X}$, where $W_{X}$ is given by the restriction of $W$ to $\mathbb{R}\times \mathbb{R} \times\Pi^{1} \left( \mathcal{B} , \mathcal{B}\right)_{X}^{X}$,
$$ W_{X}: \mathbb{R}\times \mathbb{R} \times\Pi^{1} \left( \mathcal{B} , \mathcal{B}\right)_{X}^{X} \rightarrow V$$
In other words, the $X-$material distribution of $\mathcal{C}$ is generated by the left-invariant vector fields $\Theta$ on $\mathbb{R}\times \mathbb{R} \times\Pi^{1} \left( \mathcal{B} , \mathcal{B}\right)_{X}^{X}$ such that
\begin{equation}\label{tmatgrou1.second4}
    T W_{X} \left( \Theta \right) = 0.
\end{equation}
Let $\Theta$ be a left-invariant vector field on $\mathbb{R}\times \mathbb{R} \times \Pi^{1} \left( \mathcal{B} , \mathcal{B}\right)_{X}^{X}$. Then,

\begin{equation}
    \Theta  \left(t,s, y^{i}_{j}\right)   =   \lambda\dfrac{\partial}{\partial  t } +  y^{i}_{l} \Theta^{l}_{j}\dfrac{\partial}{\partial y^{i}_{j} },
\end{equation}
respect to a local system of coordinates $\left(t,s, y^{i}_{j}\right)$ on $ \mathbb{R}\times \mathbb{R} \times \Pi^{1} \left( \mathcal{U} , \mathcal{U}\right)_{X}^{X}$ with $\mathcal{U}$ an open subset of $\mathcal{B}$ with $X \in \mathcal{U}$. Then, $\Theta$ is an admissible vector field for the couple $\left( \Phi \left( \mathcal{V} \right) , \Omega_{X} \left( \mathbb{R} \right)\right)$ if, and only if, the following equations hold,
\begin{equation}\label{Xmatgroup323456}
    \lambda\dfrac{\partial W_{X}}{\partial  t } + y^{i}_{l} \Theta^{l}_{j}\dfrac{\partial   W_{X}}{\partial  y^{i}_{j} } = 0
\end{equation}
Observe that, here $\lambda$ and $\Theta^{i}_{j}$ are function depending on $t$. This equation is called the \textit{evolution equation at $X$} \cite{EvoEqu}.\\
The base-characteristic distribution $A \Omega_{X} \left( \mathbb{R}\right)^{\sharp}$ (see theorem \ref{10.24}) will be called \textit{$X-$body-material distribution}. The foliations associated to the $X-$material distribution and the $X-$body-material distribution will be called \textit{$X-$material foliation} and \textit{$X-$body-material foliation}, and they will be denoted by $\overline{\mathcal{F}}_{X}$, respectively. It is important do not confuse $\overline{\mathcal{F}}_{X}$ (resp. $\mathcal{F}_{X}$), the $X-$material foliation (resp. $X-$body-material foliation), with $\overline{\mathcal{F}}\left(\epsilon\left( X\right)\right)$ (resp. $\mathcal{F}\left( X\right)$), the leaf at $\epsilon\left( X\right)$ (resp. the leaf at $X$) of the foliation $\overline{\mathcal{F}}$ (res. $\mathcal{F}$).}\\

\section{Morphogenesis}\label{section5}

As opposed to the uniformity in the spatial case \cite{VMMDME}, arise new material properties associated to the evolution of the body. In particular, the temporal counterpart of uniformity is a specific case of evolution of the material called \textit{remodeling}.

\begin{definition}\label{1.17.2.se}
\rm
Let $\mathcal{C}$ be a body-time manifold. A material particle $X \in \mathcal{B}$ is presenting a \textit{remodeling} when it is connected with all the instants by a material isomorphism, i.e., all the points at $\mathbb{R} \times \{X\}$ are connected by material isomorphisms. $\mathcal{C}$ is presenting a \textit{remodeling} when all the material points are presenting a remodeling. \textit{Growth} and \textit{resorption} are given by a remodeling with volume increase or volume decrease of the material body $\mathcal{B}$. 
\end{definition}
Intuitively, a material evolution presents a remodeling when the constitutive properties of the material does not change with the time. This kind of evolution may be found in biological tissues \cite{RODRIGUEZ1994455}. Wolff's law of trabecular architecture of bones (see for instance \cite{TURNER19921}) is a relevant example. Here, trabeculae are assumed to change their orientation following the principal direction of stress. It is important to note that the fact of that the material body remains materially isomorphic with the time does not preclude the possibility of adding (growth) or removing (resorption) material, as long as the material added is \textit{of the same type}. It is easy to realize that a particle \textit{$X$ is presenting a remodeling if, and only if, the $X-$material groupoid is transitive} \cite{EvoEqu}. On the other hand, \textit{$\mathcal{C}$ is presenting a remodeling if, and only if, the material groupoid is transitive} \cite{CharFol}.
\begin{definition}\label{1.17.2}
\rm
Let $\mathcal{C}$ be a body-time manifold. A material particle $X \in \mathcal{B}$ is presenting a \textit{aging} when it is not presenting a remodeling, i.e., not all the instants are connected by a material isomorphism. $\mathcal{C}$ is a \textit{process of aging} if it is not a process of remodeling.
\end{definition}
Clearly, if the material response is not preserved along the time via material isomorphism, the constitutive properties are changing with the time. Altough it is something natural, a proper definition of \textit{smooth aging} was not obtained until now \cite{EvoEqu,CharFol}.

\begin{proposition}[\cite{EvoEqu}]\label{auxprop4342}
Let $\mathcal{C}$ be a body-time manifold. A material particle $X \in \mathcal{B}$ is presenting an aging if, and only if, the $X-$material groupoid $\Omega_{X}\left(  \mathbb{R}\right)$ is not transitive. $\mathcal{C}$ is presenting an aging if, and only if, for some material point $X$, the $X-$material groupoid $\Omega_{X}\left(  \mathbb{R}\right)$ is not transitive.
\end{proposition}

In \cite{EvoEqu, CharFol}, the authors use the corresponding material distributions and foliations to presents several results characterizing the different types of remodeling and aging. Here, however, we are interested in a particular case of evolution: \textit{morphogenesis}.\\\\
\noindent{Let us start remembering a classical notion in theory of groups. Let $G$ be a group. A subgroup $N$ of $G$ is said to be \textit{normal}, and denoted by $N \trianglelefteq G$, if it is invariant under conjugation, i.e.,}
\begin{equation}\label{23.2323}
    N = g \cdot N \cdot g^{-1},
\end{equation}
for all $g \in G$. Let us consider a general subgroup $H$ of $G$. Then, the \textit{normalizer} of $H$ in $G$, denoted by $\mathcal{N} \left( H \right)$, is defined as the family of elements of $G$ in such a way that $H$ is invariant under their conjugations, i.e.,
\begin{equation}
    \mathcal{N} \left( H \right) := \{ g \in G \ / \ H = g \cdot H \cdot g^{-1} \}
\end{equation}
In fact, $\mathcal{N} \left( H \right)$ is the largest subgroup of $G$ in which $H$ is a normal subgroup.\\
In order to work deal with the concept of \textit{morphogenesis}, we will need to extent this notion to groupoids. This generalization may be found in \cite{CEDRIC12} to study uniformity and homogeneity in \textit{functional graded materials} (FGM). Here, we will develop a previous mathematical study of this extension, which is necessary for our research.
\begin{definition}\label{normalizer76465}
\rm
Let $\Gamma \rightrightarrows M$ be a groupoid. A subgroupoid $\overline{\Gamma} \rightrightarrows N$ of $\Gamma$ is said to be \textit{normal} in $\Gamma$ if it is invariant under conjugation, i.e.,
$$   g \cdot h \cdot g^{-1} \in \overline{\Gamma},$$
for all $g \in \Gamma$ and $h \in \overline{\Gamma}$ such that $\beta \left( h \right) = \alpha \left( g \right) = \alpha \left(h \right)$.
\end{definition}
Notice that, in the case of groupoid, an equation like Eq. (\ref{23.2323}) does not make sense because, in general, we cannot compose an element of $\Gamma$ with all the elements at $\overline{\Gamma}$. In general, one could think that the imposition of being a normal subgroupoid may be reduced to the isotropy groups in such a way that \textit{a subgroupoid of a groupoid is normal if, and only if, all the isotropy groups are normal in the correspondent isotropy groups of the groupoid}. Nevertheless, in general, condition of being a normal subgroupoid is more restrictive than the property of that all the isotropy groups are normal subgroups of the correspondent isotropy groups.\\
An easy counterexample is the following: consider the groupoid $\Gamma \rightrightarrows M$ such that  $\Gamma =  M \times M \times Gl \left( 3 , \mathbb{R}\right)$ and $M:=\{x,y\}$ (see example \ref{truvialG}). Here, we may consider the subgroupoid $\overline{\Gamma} \rightrightarrows M$ characterized by:
\begin{itemize}
    \item $\overline{\Gamma}_{x}^{x} =  \{x\}\times \{x\} \times Gl \left( 3 , \mathbb{R}\right)$
    \item $\overline{\Gamma}_{y}^{y} =  \{y\}\times \{y\} \times \{Id\}$
    \item $\overline{\Gamma}_{x}^{y} = \emptyset$
\end{itemize}
Then, obviously, all the isotropy groups of $\overline{\Gamma}$ are normal in the correspondent isotropy groups of $\Gamma$. However, in general, for an element $F$ of $\overline{\Gamma}_{x}^{x} =  \{x\}\times \{x\} \times Gl \left( 3 , \mathbb{R}\right)$ and an element $H$ of $\Gamma_{x}^{y} =  \{y\}\times \{y\} \times Gl \left( 3 , \mathbb{R}\right)$, it does not satisfy that
$$ H \cdot F \cdot H^{-1} \in \overline{\Gamma}_{y}^{y} = \{y\}\times \{y\} \times \{Id\}.$$
So, $\overline{\Gamma}$ is not a normal subgroupoid of $\Gamma$.
\begin{proposition}\label{propositionaux1}
Let $\Gamma \rightrightarrows M$ be a groupoid. A transitive subgroupoid $\overline{\Gamma} \rightrightarrows N$ of $\Gamma$ is normal in $\Gamma$ if, and only if, all the isotropy groups are normal in the correspondent isotropy groups of $\Gamma$.
\begin{proof}
Assume that all isotropy groups of $\overline{\Gamma}$ are normal in the correspondent isotropy groups of $\Gamma$. Then, for all $x \in M$ we have
$$ \overline{\Gamma}_{x}^{x} \trianglelefteq \Gamma_{x}^{x}, \ \forall x \in N.$$
Let $g \in \Gamma$ and $h \in \overline{\Gamma}$ such that, $\beta \left( h \right) = \alpha \left( g \right) = \alpha \left(h \right)$. Then, by transitivity, consider $l \in \overline{\Gamma}$ such that $  \alpha\left( l \right) = \alpha\left( h \right) = \beta \left( h \right) = \alpha \left( g \right) $ y $\beta \left(  l\right) = \beta \left(  g\right) $. Then,
$$  g \cdot l^{-1} \in \Gamma_{\beta \left( g \right)}^{\beta \left( g \right)} \ \ \ \ \ \ , \ \ \ \ \ \ l \cdot h \cdot l^{-1} \in \overline{\Gamma}_{\beta \left( g \right)}^{\beta \left( g \right)}$$
Therefore,
$$ g \cdot h \cdot g^{-1} = \left( g \cdot l^{-1} \right) \cdot \left( l \cdot h \cdot l^{-1}  \right) \cdot \left( g \cdot l^{-1} \right)^{-1} \in \overline{\Gamma}_{\beta \left( g \right)}^{\beta \left( g \right)}$$
\end{proof}
\end{proposition}
So, we have a wide family of subgroupoids in which the imposition of being normal may be reduced to the corresponding imposition on the isotropy groups.
\begin{definition}
\rm
Let $\Gamma \rightrightarrows M$ be a groupoid and $\overline{\Gamma} \rightrightarrows N$ be a subgroupoid of $\Gamma$. The \textit{normalizoid of $\overline{\Gamma}$ in $\Gamma$} is given by
\begin{equation}
    \mathcal{N} \left( \overline{\Gamma} \right) := \{ g \in \Gamma \ /  g \cdot h \cdot g^{-1} \in \overline{\Gamma},  \ \forall h \in \overline{\Gamma}, \  \beta \left( h \right) = \alpha \left( g \right) = \alpha \left( h \right)\}
\end{equation}
\end{definition}

\begin{proposition}\label{2.2}
Let $\Gamma \rightrightarrows M$ be a groupoid and $\overline{\Gamma} \rightrightarrows N$ be a subgroupoid of $\Gamma$. The normalizoid $\mathcal{N} \left( \overline{\Gamma} \right)$ of $\overline{\Gamma}$ in $\Gamma$ has a structure of subgroupoid of $\Gamma$ over $N$. In fact, $\mathcal{N} \left( \overline{\Gamma} \right)$ is the largest subgroupoid of $\Gamma$ satisfying that $\overline{\Gamma}$ is a normal subgroupoid of $\mathcal{N} \left( \overline{\Gamma} \right)$.
\begin{proof}
We only have to check that the composition is closed in $\mathcal{N} \left( \overline{\Gamma} \right)$. $\mathcal{N} \left( \overline{\Gamma} \right)$ is the largest subgroupoid of $\Gamma$ satisfying that $\overline{\Gamma}$ is a normal subgroupoid of $\mathcal{N} \left( \overline{\Gamma} \right)$ by construction.
\end{proof}
\end{proposition}

\begin{corollary}\label{corollaryaux1}

Let $\Gamma \rightrightarrows M$ be a groupoid and $\overline{\Gamma} \rightrightarrows N$ be a subgroupoid of $\Gamma$. The isotropy groups of the normalizoid of $\overline{\Gamma}$ in $\Gamma$ are the normalizers of the isotropy groups of $\overline{\Gamma}$ in the correspondent isotropy groups of $\Gamma$, i.e.,
$$\mathcal{N} \left( \overline{\Gamma} \right)_{x}^{x}= \mathcal{N} \left( \overline{\Gamma}_{x}^{x}\right),$$
for all $x \in N$.
\begin{proof}
\begin{small}

\begin{eqnarray*}
    \mathcal{N} \left( \overline{\Gamma} \right)_{x}^{x} &=& \{ g \in \Gamma \ /  g \cdot h \cdot g^{-1} \in \overline{\Gamma},  \ \forall h \in \overline{\Gamma}, \  \beta \left( h \right) = \alpha \left( g \right) = \alpha \left( h \right)=x = \beta \left( g \right)\}\\
    &=& \{ g \in \Gamma_{x}^{x} \ /  g \cdot h \cdot g^{-1} \in \overline{\Gamma},  \ \forall h \in \overline{\Gamma}, \  \beta \left( h \right) = \alpha \left( g \right) = \alpha \left( h \right), \}\\
    &=& \{ g \in \Gamma_{x}^{x} \ /  g \cdot h \cdot g^{-1} \in \overline{\Gamma},  \ \forall h \in \overline{\Gamma}^{x}_{x}, \}\\
    &=& \mathcal{N} \left( \overline{\Gamma}_{x}^{x}\right)
\end{eqnarray*}
\end{small}
\end{proof}
\end{corollary}
Of course, a deeper study of the normalizoid as an abstract structure in the theory of groupoid has a great interest from a mathematical point of view. Nevertheless, this could distract the reader from the main goal of this paper. So, we will leave the mathematical study of this notion here and, from now on, we will be focused on its application to materials.\\\\

\noindent{The \textit{extended material groupoid of $\mathcal{C}$}, denoted by $\mathcal{N} \left( \mathcal{C}\right)$, is defined as the normalizoid of the material groupoid of $\mathcal{C}$ as a subgroupoid of $\Phi \left( \mathcal{V}\right)$, i.e.,
$$ \mathcal{N} \left( \mathcal{C}\right) = \mathcal{N} \left(\Omega \left( \mathcal{C}\right)\right).$$}
Thus, more explicitly,\\\\
\noindent{\textit{The elements of $\mathcal{N} \left( \mathcal{C}\right)$ are the triples $\left(t,s , j_{X,Y}^{1} \psi\right)$, with $\psi$ a local automorphism on $\mathcal{B}$, such that, $\left(s,s , j_{Y,Y}^{1} \left(\psi \circ \phi \circ \psi^{-1}\right)\right)$ is a material isomorphism, for all material symmetry $\left( t , t , j_{X,X}^{1} \phi \right)$ on $\left(t, X \right)$.}}\\\\

\noindent{Notice that, by proposition \ref{2.2}, \textit{all the material isomorphisms are contained in $\mathcal{N} \left( \mathcal{C} \right)$}. In fact, $\mathcal{N} \left( \mathcal{C} \right)$ is the largest subgroupoid of $\Phi \left( \mathcal{V}\right)$ satisfying that $\Omega \left( \mathcal{C} \right)$ is a normal subgroupoid of $\mathcal{N} \left( \mathcal{C} \right)$.\\
Analogously we define the \textit{$X$-extended material groupoid of $\mathcal{C}$}, denoted by $\mathcal{N}_{X} \left( \mathbb{R}\right)$, as the normalizoid of the $X-$material groupoid of $\mathcal{C}$. Thus, in general, we have the following short exact sequences,}

$$\Omega \left( \mathcal{C} \right) \leq \mathcal{N} \left( \mathcal{C} \right) \leq \Phi \left( \mathcal{V} \right)$$
$$\Omega_{X} \left( \mathbb{R} \right) \leq \mathcal{N}_{X} \left( \mathbb{R} \right) \leq \mathbb{R}\times \mathbb{R} \times\Pi^{1} \left( \mathcal{B} , \mathcal{B}\right)_{X}^{X}$$

\vspace{0.7cm}

\noindent{Let us now come back to material evolution. An example of aging process may be the weakening of the stiffness of a mineralized bone due to hormonal deficiencies. However, the resulting bone could still be isotropic. Then, there has not been a qualitative change of the constitutive properties. This fact, leads to identify a particular kind of aging: \textit{morphogenesis}.}

\begin{definition}\label{defintiowmorphsf3r}
\rm
Let $\mathcal{C}$ be a body-time manifold. A material point $X$ is said to be undergone a \textit{process of evolution without morphogenesis} when its symmetry groups at all the instants are conjugated. $\mathcal{C}$ is said to be undergone a \textit{process of evolution without morphogenesis} when all the points are undergone process of evolution without morphogenesis.
\end{definition}
Let us clarify the definition. Fix a material particle $X$ of the body. Remember that, for each instant $t$, the symmetry group $G \left( t,X \right)$ is given by all the material isomorphisms from $\left( t,X \right)$ to $\left( t,X \right)$, i.e.,
$$G \left( t,X \right) = \Omega \left( \mathcal{C} \right)_{\left(t,X\right)}^{\left( t,X\right)}  = \Omega_{X}\left(   \mathbb{R} \right)_{t}^{t}$$
Then, $X$ is presenting a process of evolution without morphogenesis iff for any two instant $t$ and $s$, there exists a triple $\left(s,t, j_{X,X}^{1}\phi \right) \in \Phi  \left( \mathcal{V} \right)$ such that
\begin{equation}\label{2.3}
G \left( t,X \right) =  \left(s,t, j_{X,X}^{1}\phi \right) \cdot G \left( s,X \right) \cdot \left(t,s, j_{X,X}^{1}\phi^{-1} \right)
\end{equation}

Suppose that $\mathcal{C}$ is a remodeling process. Then, connecting via material isomorphisms, all the symmetry groups are conjugated. Hence,\textit{ remodeling is an example of process of evolution without morphogenesis}. In other words, as it is natural, any process of morphogenesis is a particular case of aging.\\
Nevertheless, not all the processes of evolution without morphogenesis are remodelings. In particular, we have a process of aging without morphogenesis when not all the symmetry groups may be conjugated by material isomorphisms, but for arbitrary elements of $\Phi \left( \mathcal{V} \right)$.\\
Thus, a process of morphogenesis entails a \textit{breakdown of symmetry}, a sudden change in the material symmetry type. A state of instability could provoke that the deviations from spherical symmetry of an embryo (in its spherical blastula stage) tend to grow reaching a new equilibrium in which the symmetries have changed (see \cite{TURING232}).\\
Many others king of morphogenesis may be happen in the realm of solids bodies. In \cite{EPSTEIN201572} the author study how could change the \textit{type of symmetry} from isotropy to transverse isotropy or orthotropy.\\
\begin{proposition}
Let $\mathcal{C}$ be a body-time manifold. A material point $X$ is undergone a process of evolution without morphogenesis if, and only if, the $X-$extended material groupoid $\mathcal{N}_{X}\left( \mathbb{R} \right)$ is transitive. $\mathcal{C}$ is presenting a process of evolution without morphogenesis if, and only if, the extended material groupoid $\mathcal{N}\left( \mathcal{C} \right)$ is transitive.
\begin{proof}
Notice that, in particular, a material point $X$ presents a process of evolution without morphogenesis if, and only if, for each two instants $t$ and $s$, there exists an element $g = \left( s, t , j_{X,X}^{1}\phi   \right)$ of $\Phi \left( \mathcal{V}\right)$ from $\left( s,X \right)$ to $\left( t , X \right)$ such that
\begin{equation}\label{Moreandmoreequations24}
\Omega_{X}\left(\mathbb{R}\right)_{t}^{t} = g \cdot \Omega_{X}\left(\mathbb{R}\right)_{s}^{s} \cdot g^{-1},
\end{equation}
where $\Omega_{X}\left(\mathbb{R}\right)_{r}^{r}$ is the isotropy group at $\left( r, X \right)$ for all $s$. In other words, $X$ presents a process of evolution without morphogenesis if, and only if, for each two instants $t$ and $s$, there exists an element $g $ of the $X-$extended material groupoid $\mathcal{N}_{X}\left( \mathbb{R} \right)$ from $\left( t,X \right)$ to $\left( s , X \right)$ or, in other words, if and only if $\mathcal{N}_{X}\left( \mathbb{R} \right)$ is transitive. The proof for $\mathcal{N}\left( \mathcal{C} \right)$ is analogous.
\end{proof}
\end{proposition}
\noindent{So, the normalizoid of the $X-$material groupoids works as a tool to study the morphogenesis of a particle, while the normalizoid of the material groupoid works to study the morphogenesis of the whole body, both via the property of transitivity. However, in general, it not easy to check the transitivity of a groupoid. To solve this problem, we will use the associated characteristic distribution}.\\

\noindent{Now, we will consider the associated characteristic distribution $A \mathcal{N} \left( \mathcal{C} \right)^{T}$ to the extended material groupoid of $\mathcal{C}$, which will be called \textit{extended material distribution}.} The base-characteristic distribution $A \mathcal{N} \left( \mathcal{C} \right)^{\sharp}$ will be called \textit{extended body-material distribution}.\\
The foliations associated to the extended material distribution and the extended body-material distribution will be called \textit{extended material foliation} and \textit{extended body-material foliation} and they will be denoted by $\mathcal{NF}$ and $\overline{\mathcal{NF}}$, respectively.\\
Then, as a consequence of theorem \ref{10.20} and corollary \ref{10.39}, we have that,
\begin{theorem}\label{Th1morphogenesis}
For each $\left(X, t \right) \in \mathcal{C}$, there exists a transitive Lie subgroupoid $\mathcal{N}  \left( X, t \right) $ of $\Phi \left( \mathcal{V}\right)$ with base $\mathcal{NF} \left( X, t \right) $, which is totally contained in $\mathcal{N} \left( \mathcal{C} \right)$. Furthermore, $\mathcal{NF}$ is the coarsest foliation of $\mathcal{C}$ satisfying this property.
\end{theorem}

Notice that, over each leaf $\mathcal{NF} \left( X, t \right) $, there exists a \textit{transitive Lie} subgroupoid $\mathcal{N}  \left( X, t \right) $ contained in $\mathcal{N} \left( \mathcal{C} \right)$. This fact is interpreted as each leaf is a differentiable process of morphogenesis. Thus, we would like to highlight that this mathematical result has a very intuitive meaning:\\

\framebox{%
  \begin{minipage}{1\linewidth}
\textit{The body-time manifold $\mathcal{C}$ may be canonically divided, in a maximal way, into submanifolds (the foliation $\mathcal{NF}$) in such a way that all the leaves are differentiable processes of evolution without morphogenesis.}
\end{minipage}
}

\vspace{0.25cm}

As we commented, to study the morphogenesis is not an easy problem using unically the transitivity of the extended material groupoid. Then, following result present a ``\textit{computable}'' way to study morphogenesis via theorem \ref{Th1morphogenesis}.
\begin{theorem}\label{extendedmaterialgroupoids24456}
Let $\mathcal{C}$ be a body-time manifold and $\mathcal{N} \left( \mathcal{C} \right)$ its extended material groupoid. The extended material distribution $A \mathcal{N} \left( \mathcal{C} \right)^{T}$ is pointwise generated by the left-invariant vector fields $\Theta$ on $\Phi \left( \mathcal{V} \right)$ satisfying that
\begin{equation}\label{normalizoid2331146}
 TW \left( \left[ \Theta , \Lambda \right] \right) = 0,
 \end{equation}
for all left invariant admissible vector field $\Lambda$ for the couple $ \left( \Phi \left( \mathcal{V}\right),  \Omega \left( \mathcal{C} \right) \right)$ which is tangent to the $\alpha-$fibres.
\begin{proof}
 Let $\Theta \in \mathfrak X_{loc} \left( \Phi \left( \mathcal{V} \right) \right)$ be a (local) left-invariant vector field on $\Phi \left( \mathcal{V} \right)$ whose (local) flow is denoted by $\varphi^{\Theta}_{t}$. Let us also denote by $\Theta^{\sharp} \in \mathfrak X_{loc} \left(  \mathcal{C} \right)$ the (local) the projection of $\Theta$ on the base-characteristic distribution of $\Phi \left( \mathcal{V}\right)$, i.e.,
 
 $$ \Theta^{\sharp} = T \alpha \circ \Theta \circ \epsilon$$
 The local flow of $\Theta^{\sharp}$ will be denoted by $\varphi_{t}^{\Theta^{\sharp}}$. Notice that, we have $\varphi_{t}^{\Theta^{\sharp}} = \alpha \circ \varphi_{t}^{\Theta} \circ \epsilon$.\\

\noindent{Then, by definition, $\Theta$ is admissible for the couple $\left( \Phi \left( \mathcal{V} \right) , \mathcal{N} \left( \mathcal{C} \right) \right)$ if, and only if, $\varphi_{t}^{\Theta}$ is completely contained in $\mathcal{N} \left( \mathcal{C} \right)$ at the identities. In other words,}
\begin{equation}\label{auxiliar123123eq}
W \left( F   \cdot \varphi_{t}^{\Theta} \left( \epsilon \left( r,X \right) \right) \cdot h \cdot \varphi_{-t}^{\Theta} \left( \epsilon \left( \varphi_{t}^{\Theta^{\sharp}} \left( r,X\right) \right) \right)  \right) = W \left( F \right),
\end{equation}
for all $\left(r, X \right) \in \mathcal{C}$, $F \in \Phi \left( \mathcal{V} \right)$ and $h \in \Omega \left( \mathcal{C} \right)$ with $\alpha \left( F \right) = \left(r, X\right)$ and $\alpha \left( h \right) = \beta \left( h \right) = \varphi_{t}^{\Theta^{\sharp}} \left( r,X\right)$.\\
By simplicity, the identity at $\left( r , X \right)$ in $\Phi \left( \mathcal{V} \right)$, $\left( r,r ,j_{X,X}^{1}Id_{T_{X}\mathcal{B}}   \right)$ with $Id_{T_{X}\mathcal{B}}$ the identity map on $T_{X}\mathcal{B}$, is denoted by $\epsilon \left( r,X \right) $.\\
Let us now consider a left invariant vector field $\Lambda$ on $\Phi \left( \mathcal{V} \right)$ tangent to the $\alpha-$fibres whose (local) flow is denoted by $\varphi^{\Lambda}_{t}$. Then, it satisfies that
\begin{eqnarray*}
& & W \left( F   \cdot \varphi_{t}^{\Theta} \left( \epsilon \left( r,X \right) \right) \cdot \varphi^{\Lambda}_{s} \left(  \epsilon \left( \varphi_{t}^{\Theta^{\sharp}} \left( r,X\right) \right)   \right) \cdot \varphi_{-t}^{\Theta} \left( \epsilon \left( \varphi_{t}^{\Theta^{\sharp}} \left( r,X\right) \right) \right)  \right) \\
 & & =  W \left( \varphi_{-t}^{\Theta} \left(\varphi^{\Lambda}_{s} \left(  \varphi_{t}^{\Theta} \left( F \right)\right) \right) \right)\\
 & &  = W \left( F \right)
\end{eqnarray*}
Notice that, since it is tangent to the $\alpha-$fibres, $\varphi^{\Lambda}_{t}$ have the same projection by $\beta$ and $\alpha$ (so, the above composition makes sense). Then, derivating with respect to the variable $s$ at $0$, we get
\begin{equation}
    T_{F} W \left( {\varphi^{\Theta}_{-t}}_{*} \Lambda \left( F \right) \right) = 0,
\end{equation}
where ${\varphi^{\Theta}_{-t}}_{*}\Lambda $ is the pushforward  of $\Lambda$ by $\phi^{\Theta}_{-t}$. Therefore, derivating with respect to the variable $t$ at $0$, we have that
\begin{equation}\label{auxeqnormalizoid23135}
    T_{F} W \left( \left[ \Theta , \Lambda \right] \left( F \right)\right) = 0.
\end{equation}
Conversely, assume that Eq. (\ref{auxeqnormalizoid23135}) is satisfied for all left invariant admissible vector field $\Lambda$, tangent to the $\alpha$ fibres, for the material distribution $A \Omega \left( \mathcal{C} \right)^{T}$. Using the same notation for the (local) flows, notice that, in general
\small{
\begin{eqnarray*}
0 &= & \dfrac{\partial }{\partial t_{\vert 0}}\left(\dfrac{\partial }{\partial s_{\vert 0}}\left(W \left( \varphi_{-t}^{\Theta} \left(\varphi^{\Lambda}_{s} \left(  \varphi_{t}^{\Theta} \left( F \right)\right) \right) \right) \right) \right)\\
 &=& - \dfrac{\partial }{\partial t_{\vert 0}}\left(\dfrac{\partial }{\partial s_{\vert 0}}\left(W \left( \varphi_{-t}^{\Lambda} \left(\varphi^{\Theta}_{s} \left(  \varphi_{t}^{\Lambda} \left( F \right)\right) \right)\right)  \right) \right)\\
 &=& - \dfrac{\partial }{\partial t_{\vert 0}}\left(\dfrac{\partial }{\partial s_{\vert 0}}\left(W \left( F   \cdot \varphi_{t}^{\Lambda} \left( \epsilon \left( r,X \right) \right) \cdot \varphi^{\Theta}_{s} \left(  \epsilon  \left( r,X\right)  \right) \cdot \varphi_{-t}^{\Lambda} \left( \epsilon  \left( \varphi_{s}^{\Theta^{\sharp}} \left( r,X\right)\right) \right) \right)  \right) \right)\\
 &=&- \dfrac{\partial }{\partial t_{\vert 0}}\left(\dfrac{\partial }{\partial s_{\vert 0}}\left(W \left( F   \cdot \varphi_{t}^{\Lambda} \left( \epsilon \left( r,X \right) \right) \cdot \varphi^{\Theta}_{s} \left(  \epsilon  \left( r,X\right)  \right)  \right)  \right) \right)\\
\end{eqnarray*}
}
\noindent{for all $F$. Here, we are using that $\varphi^{\Lambda}_{t}$ is totally contained in $\Omega \left( \mathcal{C} \right)$. Then, by changing $F$ by $G=F   \cdot \varphi_{l}^{\Lambda} \left( \epsilon \left( r,X \right) \right) $ we have that}
\small{
\begin{eqnarray*}
& & \dfrac{\partial }{\partial t_{\vert l}}\left(\dfrac{\partial }{\partial s_{\vert 0}}\left(W \left( \varphi_{-t}^{\Theta} \left(\varphi^{\Lambda}_{s} \left(  \varphi_{t}^{\Theta} \left( F \right)\right) \right) \right) \right) \right)\\
 &=& - \dfrac{\partial }{\partial t_{\vert l}}\left(\dfrac{\partial }{\partial s_{\vert 0}}\left(W \left( F   \cdot \varphi_{t}^{\Lambda} \left( \epsilon \left( r,X \right) \right) \cdot \varphi^{\Theta}_{s} \left(  \epsilon  \left( r,X\right)  \right) \right)  \right) \right)\\
 &=& - \dfrac{\partial }{\partial t_{\vert 0}}\left(\dfrac{\partial }{\partial s_{\vert 0}}\left(W \left(G \cdot \varphi_{t}^{\Lambda} \left( \epsilon \left( r,X \right) \right) \cdot \varphi^{\Theta}_{s} \left(  \epsilon  \left( r,X\right)  \right) \right)\right)\right)\\
 &=& 0
\end{eqnarray*}
}

\noindent{Therefore,} 

$$\dfrac{\partial }{\partial s_{\vert 0}}\left(W \left( \varphi_{-t}^{\Theta} \left(\varphi^{\Lambda}_{s} \left(  \varphi_{t}^{\Theta} \left( F \right)\right) \right) \right) \right) = 0.$$

\noindent{On the other hand,}
\begin{scriptsize}

\begin{eqnarray*}
& & \dfrac{\partial }{\partial s_{\vert l}}\left(W \left( \varphi_{-t}^{\Theta} \left(\varphi^{\Lambda}_{s} \left(  \varphi_{t}^{\Theta} \left( F \right)\right) \right) \right) \right) \\
&=&  \dfrac{\partial }{\partial s_{\vert l}}\left(W \left( F   \cdot \varphi_{t}^{\Theta} \left( \epsilon \left( r,X \right) \right) \cdot \varphi^{\Lambda}_{s} \left(  \epsilon \left( \varphi_{t}^{\Theta^{\sharp}} \left( r,X\right) \right)   \right) \cdot \varphi_{-t}^{\Theta} \left( \epsilon \left( \varphi_{t}^{\Theta^{\sharp}} \left( r,X\right) \right) \right)  \right)\right)\\
&=&  \dfrac{\partial }{\partial s_{\vert 0}}\left(W \left( H \cdot \varphi_{t}^{\Theta} \left( \epsilon \left( r,X \right) \right) \cdot \varphi^{\Lambda}_{s} \left(  \epsilon \left( \varphi_{t}^{\Theta^{\sharp}} \left( r,X\right) \right)   \right) \cdot \varphi_{-t}^{\Theta} \left( \epsilon \left( \varphi_{t}^{\Theta^{\sharp}} \left( r,X\right) \right) \right)  \right)\right)\\
&=&\dfrac{\partial }{\partial s_{\vert 0}}\left(W \left( \varphi_{-t}^{\Theta} \left(\varphi^{\Lambda}_{s} \left(  \varphi_{t}^{\Theta} \left( H \right)\right) \right) \right) \right) \\
&=& 0
\end{eqnarray*}
\end{scriptsize}
Where, by simplicity, we are considering the following change of notation,

$$H = F   \cdot \varphi_{t}^{\Theta} \left( \epsilon \left( r,X \right) \right) \cdot \varphi^{\Lambda}_{l} \left(  \epsilon \left( \varphi_{t}^{\Theta^{\sharp}} \left( r,X\right) \right)   \right) \cdot \varphi_{-t}^{\Theta} \left( \epsilon \left( \varphi_{t}^{\Theta^{\sharp}} \left( r,X\right) \right) \right)$$
There, we have already proved that,
$$W \left( \varphi_{-t}^{\Theta} \left(\varphi^{\Lambda}_{s} \left(  \varphi_{t}^{\Theta} \left( F \right)\right) \right) \right) \equiv Cte.$$
So, evaluating at $t=s=0$,
$$W \left( \varphi_{-t}^{\Theta} \left(\varphi^{\Lambda}_{s} \left(  \varphi_{t}^{\Theta} \left( F \right)\right) \right) \right) \equiv W \left( F \right), \ \forall F.$$
Hence, Eq. (\ref{auxiliar123123eq}) is satisfied.

\end{proof}
\end{theorem}
\vspace{1cm}
As we promised, this result will give us a computational way of studying the morphogenesis property. Let be a local system of coordinates $\left( t,s, x^{i} , y^{j}, y^{i}_{j}\right) $ on $\Phi \left( \mathcal{V} \right)$. Consider a left-invariant vector field $ \Theta$ on $\Phi \left( \mathcal{V} \right)$ and a left-invariant vector field $\Lambda$, tangent to the $\alpha$ fibres, on $\Phi \left( \mathcal{V} \right)$,
\small{
\begin{equation}
    \Theta  \left(t,s,x^{i} , y^{j}, y^{i}_{j}\right)   = \lambda \dfrac{\partial}{\partial  t}   +   \Theta^{i}\dfrac{\partial}{\partial  x^{i} } + y^{i}_{l} \Theta^{l}_{j}\dfrac{\partial}{\partial y^{i}_{j} }, \ \ \ \ \ \Lambda  \left(t,s,x^{i} , y^{j}, y^{i}_{j}\right)   =  y^{i}_{l} \Lambda^{l}_{j}\dfrac{\partial}{\partial y^{i}_{j} }
\end{equation}
}
Therefore,
$$\left[  \Theta   , \Lambda     \right]    =   y^{i}_{l}\left( \Theta^{l}_{r}\Lambda^{r}_{j}  +  -    \Lambda^{l}_{r}\Theta^{r}_{j} + \Theta^{k}\dfrac{\partial\Lambda^{l}_{j}}{\partial x^{k}}    +  \lambda \dfrac{\partial \Lambda_{j}^{l}}{\partial t}    \right) \dfrac{\partial}{\partial y^{i}_{j} } $$
Then, by using the evolution equation (\ref{matgrou2}), the equation to be satisfied is,
\begin{equation}\label{first34eqanother455}
y^{i}_{l}\left( \Theta^{l}_{r}\Lambda^{r}_{j}  + \Theta^{k}\dfrac{\partial\Lambda^{l}_{j}}{\partial x^{k}}    +  \lambda \dfrac{\partial \Lambda_{j}^{l}}{\partial t}    -    \Lambda^{l}_{r}\Theta^{r}_{j} \right) \dfrac{\partial   W}{\partial y^{i}_{j} } = 0
\end{equation}
for all matrix function $\left(\Lambda^{i}_{j}\right)$ fulfilling the equation,
\begin{equation}\label{anothereqatuoidasd23}
y^{i}_{l} \Lambda^{l}_{j}\dfrac{\partial   W}{\partial  y^{i}_{j} } = 0
\end{equation}
Here the functions $\lambda$, $\Theta^{i}$, $\Theta^{j}_{i}$ and $\Lambda^{j}_{i}$ depends on the variables $t$ and $X$. In practice, it is enough to solve Eq. (\ref{first34eqanother455}) for a basis of the space of solutions of Eq. (\ref{anothereqatuoidasd23}). Thus, the extended material distribution is pointwise generated by the (local) functions $\lambda$, $\Theta^{i}$ and $\Theta^{j}_{i}$ satisfying the Eq. (\ref{first34eqanother455}) for all functions $\Lambda^{j}_{i}$ satisfying Eq. (\ref{anothereqatuoidasd23}). This equation will be called the \textbf{morphogenesis equation} by vitue of proposition \ref{morhoequat22}.\\\\

\noindent{Analogously, we will consider the associated characteristic distribution $A \mathcal{N}_{X} \left( \mathbb{R} \right)^{T}$ to the extended material groupoid of $\mathcal{C}$, which will be called \textit{$X-$extended material distribution}. The base-characteristic distribution $A \mathcal{N}_{X} \left( \mathbb{R} \right)^{\sharp}$ will be called \textit{$x$extended body-material distribution}. The foliations associated to the $x-$extended material distribution and the $x-$extended body-material distribution will be called \textit{extended material foliation} and \textit{extended body-material foliation} and they will be denoted by $\mathcal{NF}_{X}$ and $\overline{\mathcal{NF}}_{X}$, respectively.\\

\begin{theorem}\label{Th1morphogenesisX}
For each material particle $X \in \mathcal{B}$ and each instant $t$, there exists a transitive Lie subgroupoid $\mathcal{N}_{X}  \left(t \right) $ of $\mathbb{R}\times \mathbb{R} \times\Pi^{1} \left( \mathcal{B} , \mathcal{B}\right)_{X}^{X}$ with base $\mathcal{NF}_{X} \left( t \right) $, which is totally contained in $\mathcal{N}_{X} \left( \mathbb{R} \right)$. Furthermore, $\mathcal{NF}_{X}$ is the coarsest foliation of $\mathbb{R}$ satisfying this property.
\end{theorem}

In other words, we have proved the following:

\framebox{%
  \begin{minipage}{1\linewidth}
\textit{For each particle $X$, the time-line $\mathbb{R}$ may be canonically divided, in a maximal way, into submanifolds (intervals or points given by the foliation $\mathcal{NF}_{X}$) in such a way that all the leaves are differentiable processes of evolution without morphogenesis of $X$.}
\end{minipage}
}

\vspace{0.25cm}

Therefore, to calculate this foliation permits us to now, exactly, in what instants of time differentiable process of morphogenesis are being produced.

Analogously to theorem \ref{extendedmaterialgroupoids24456}, we hve the following result:}
\begin{proposition}\label{anotherprop24445}
Let $\mathcal{C}$ be a body-time manifold and $\mathcal{N}_{X} \left( \mathbb{R} \right)$ the $X-$extended material groupoid. The extended material distribution $A \mathcal{N}_{X} \left( \mathbb{R} \right)^{T}$ is pointwise generated by the left-invariant vector fields $\Theta$ on $\mathbb{R}\times \mathbb{R} \times\Pi^{1} \left( \mathcal{B} , \mathcal{B}\right)_{X}^{X}$ satisfying that
\begin{equation}\label{punctualnormalizoid2331146}
 TW_{X} \left( \left[ \Theta , \Lambda \right] \right) = 0,
 \end{equation}
for all left invariant admissible vector field $\Lambda$ for the couple $ \left(\mathbb{R}\times \mathbb{R} \times\Pi^{1} \left( \mathcal{B} , \mathcal{B}\right)_{X}^{X},  \Omega_{X} \left( \mathbb{R}\right)  \right)$ which tangent to the $\alpha-$fibres.
\begin{proof}
It may be proved analogously to proposition \ref{extendedmaterialgroupoids24456}.

\end{proof}
\end{proposition}

Let be a local system of coordinates $\left( t,s,  y^{i}_{j}\right) $ on $\mathbb{R}\times \mathbb{R} \times\Pi^{1} \left( \mathcal{B} , \mathcal{B}\right)_{X}^{X}$. Consider a left-invariant vector field $ \Theta$ on $\mathbb{R}\times \mathbb{R} \times\Pi^{1} \left( \mathcal{B} , \mathcal{B}\right)_{X}^{X}$ and a left-invariant vector field $\Lambda$ on $\mathbb{R}\times \mathbb{R} \times\Pi^{1} \left( \mathcal{B} , \mathcal{B}\right)_{X}^{X}$, that is
\begin{equation}
    \Theta  \left(t,s, y^{i}_{j}\right)   = \lambda \dfrac{\partial}{\partial  t}    + \Theta^{i}_{j}\dfrac{\partial}{\partial y^{i}_{j} }, \ \ \ \ \ \Lambda  \left(t,s, y^{i}_{j}\right)   =  \Lambda^{i}_{j}\dfrac{\partial}{\partial y^{i}_{j} }
\end{equation}
Therefore,
$$\left[  \Theta   , \Lambda     \right]    =   y^{i}_{l}\left( \Theta^{l}_{r}\Lambda^{r}_{j}      +  \lambda \dfrac{\partial \Lambda_{j}^{l}}{\partial t}    -    \Lambda^{l}_{r}\Theta^{r}_{j} \right) \dfrac{\partial}{\partial y^{i}_{j} } $$

\noindent{Hence, by taking into account the evolution equation for $X$ (\ref{tmatgrou1.second4}), the equation to be satisfied is,}
\begin{equation}\label{eqanother455}
y^{i}_{l}\left( \Theta^{l}_{r}\Lambda^{r}_{j}  +    +  \lambda \dfrac{\partial \Lambda_{j}^{l}}{\partial t}    -    \Lambda^{l}_{r}\Theta^{r}_{j} \right) \dfrac{\partial   W_{X}}{\partial y^{i}_{j} } = 0
\end{equation}
for all matrix function $\left(\Lambda^{i}_{j}\right)$ which satisfies,
\begin{equation}\label{22anothereqatuoidasd23}
y^{i}_{l} \Lambda^{l}_{j}\dfrac{\partial   W_{X}}{\partial  y^{i}_{j}} = 0
\end{equation}
It is important to highlight that, in this case, the functions $\lambda$, $\Theta^{i}$, $\Theta^{j}_{i}$ and $\Lambda^{j}_{i}$ depends on the variable $t$. Analogously to the previous case, in practice, we will only have to check the Eq. (\ref{eqanother455}), for a basis of the space of solutions of Eq. (\ref{22anothereqatuoidasd23}).\\
Again, the $X-$extended material distribution is pointwise generated by the (local) functions $\lambda$, $\Theta^{i}$ and $\Theta^{j}_{i}$ satisfying the Eq. (\ref{eqanother455}) for all functions $\Lambda^{j}_{i}$ satisfying Eq. (\ref{22anothereqatuoidasd23}). Thus, this equation will be called the \textbf{morphogenesis equation for X}.\\

The base-characteristic distribution $A \mathcal{N}_{X} \left( \mathbb{R} \right)^{\sharp}$ will be called \textit{$X-$extended body-material distribution}. The foliations associated to the $X-$extended material distribution and the $X-$extended body-material distribution will be called \textit{$X-$extended material foliation} and \textit{$X-$extended body-material foliation} and they will be denoted by $\overline{\mathcal{NF}}_{X}$ and $\mathcal{NF}_{X}$, respectively.\\\\

As last theoretical results of the paper, we will present explicitly how may we use these morphogenesis equations to study the morhogenesis of the evolution.

\begin{proposition}[Global morphogenesis at $X$]
Let be a body-time manifold $\mathcal{C}$ and a material point $X$. $X$ presents a smooth process without morphogenesis if, and only if, 
$$    dim  \left( A \mathcal{N}_{X} \left( \mathbb{R}\right)^{\sharp}_{t} \right)  =1.$$
for all instants $t$, with $A \mathcal{N}_{X} \left( \mathbb{R}\right)^{\sharp}_{t}$ the fibre of $A \mathcal{N}_{X} \left( \mathbb{R}\right)^{\sharp}$ at $t$.
\begin{proof}
We only have to observe that the imposed condition implies that $\mathcal{N}_{X} \left( \mathbb{R}\right)$ is a transitive Lie subgroupoid of $\Phi \left( \mathcal{V} \right)$.
\end{proof}
\end{proposition}

In fact, we may generalize the result as follows:
\begin{proposition}[Local morphogenesis at $X$]
Let be a body-time manifold $\mathcal{C}$, a material point $X$ and an instant $t$. $X$ presents a smooth process without morphogenesis on an interval of time around $t$ if, and only if, 
$$    dim  \left( A \mathcal{N}_{X} \left( \mathbb{R}\right)^{\sharp}_{t} \right)  =1.$$
with $A \mathcal{N}_{X} \left( \mathbb{R}\right)^{\sharp}_{t}$ the fibre of $A \mathcal{N}_{X} \left( \mathbb{R}\right)^{\sharp}$ at $t$.
\end{proposition}
Thus, to study if the evolution of a particle $X$ is not producing morphogenesis (local or global), we will have to deal with the morphogenesis equation for $X$ (\ref{eqanother455}), 
\begin{equation*}
y^{i}_{l}\left( \Theta^{l}_{r}\Lambda^{r}_{j}  +    +  \lambda \dfrac{\partial \Lambda_{j}^{l}}{\partial t}    -    \Lambda^{l}_{r}\Theta^{r}_{j} \right) \dfrac{\partial   W_{X}}{\partial y^{i}_{j} } = 0
\end{equation*}
for all functions $\Lambda^{j}_{i}$ satisfying Eq. (\ref{22anothereqatuoidasd23}),
which is the equation whose solutions generate the $X-$extended material distribution.\\

\framebox{%
  \begin{minipage}{1\linewidth}
  In particular, $X$ presents a (local) smooth process without morphogenesis if, and only if, there exists a (local) solution $\left( \lambda , \Theta^{i}_{j}\right)$
of the morphogenesis equation fo $X$ with $\lambda \neq 0$, for all $\Lambda^{i}_{j}$ satisfying Eq. (\ref{22anothereqatuoidasd23}).
\end{minipage}
}

\noindent{Notice that, to prove that a process of aging is not producing morphogenesis, we will need to deal with isomorphisms which are not in the material groupoid, i.e., isomorphisms satisfying Eq. (\ref{Moreandmoreequations24}). In particular, the differentiability of a process of aging ($\Omega_{X}\left( \mathbb{R}\right)$ or $\Omega\left( \mathcal{C} \right)$ Lie subgroupoids) does not have relation with the differentiability of the elections of the implants $g$ proving that the process is not a morphogenesis, i.e., satistying Eq. (\ref{Moreandmoreequations24}).}\\

\begin{proposition}[Global morphogenesis]\label{morhoequat22}
Let be a body-time manifold $\mathcal{C}$. $\mathcal{C}$ presents a smooth evolution process without morphogenesis if, and only if, 

\begin{itemize}
    \item[i)] $dim  \left( A \Omega \left( \mathcal{C} \right)^{T}_{\epsilon \left(\left(t,X\right)\right)}\right)  $ is constant respect to $\left(t,X \right)$\\
    \item[ii)] For some $X$, $dim  \left( A \Omega_{X} \left( \mathbb{R} \right)^{\sharp}_{t} \right)= 0 $, for some $t$.
    \item[iii)] $    dim  \left( A \mathcal{N}_{X} \left( \mathbb{R}\right)^{\sharp}_{t} \right)  =1.$
\end{itemize}
Here, $ A \Omega \left( \mathcal{C} \right)^{T}_{\epsilon \left(\left(t,X\right)\right)}$ (resp. $A \Omega_{X} \left( \mathbb{R} \right)^{\sharp}_{t}$ and  $A \mathcal{N}_{X} \left( \mathbb{R}\right)^{\sharp}_{t}$) is the fibre of $ A \Omega \left( \mathcal{C} \right)^{T}$ (resp. $A \Omega_{X} \left( \mathbb{R} \right)^{\sharp}$ and $A \mathcal{N}_{X} \left( \mathbb{R}\right)^{\sharp}_{t}$) at $\epsilon \left( \left(t,X\right)\right)$ (resp. $t$).

\end{proposition}
In this way, to prove that a material evolution present a smooth aging without morphogenesis, we will have to deal with Eq. (\ref{Eqmaterialgroupoid12}), Eq. (\ref{Xmatgroup323456}) and Eq. (\ref{first34eqanother455}).

\section{Conclusions and future work}

Using the concept of groupoid, we have developed a simple and complete geometric theory of the phenomenon of material evolution and, in particular, of the so-called morphogenesis. We have also obtained the corresponding linear equations of evolution, making our results computable.

In a future work we wish to deepen this theory, and also consider more complex situations such as in media with microstructure or composite media.

\section*{Acknowledgments}
M. de Leon and V. M. Jiménez acknowledge the partial finantial support from MICINN Grant PID2019-106715GB-C21 and the ICMAT Severo Ochoa project CEX2019-000904-S. M. de León also acknowledges MICINN’s support under grant EIN2020-112107.

\bibliographystyle{plain}

\bibliography{Library}

\end{document}